\documentclass[11pt,authoryear]{elsarticle}

\usepackage{amsmath}
\usepackage{amsfonts}
\usepackage{amssymb}
\usepackage{fancyvrb}
\usepackage{float}
\usepackage{graphicx}
\usepackage{subfig}
\usepackage[usenames,dvipsnames,svgnames,table]{xcolor}
\usepackage{algorithm}
\usepackage{algpseudocode}
\usepackage{listings}
\usepackage{tikz}
\usepackage{pgfplots}
\usepackage[margin=2.54cm]{geometry}

\usetikzlibrary{decorations.text}
\usetikzlibrary{calc}
\usetikzlibrary{external}

\newcommand{\url}[1]{{\tt \small #1}}

\newcommand{\BParm}{\mathbf{\Theta}}
\newcommand{\BHyp}{\mathbf{H}}
\newcommand{\BData}{\mathbf{D}}
\newcommand{\BPosteriorProbL}{Pr \left( \BParm \vert \BData, \BHyp \right)}
\newcommand{\BPosteriorProbS}{P \left( \BParm \vert \BData \right)}
\newcommand{\BPriorProbL}{Pr \left( \BParm \vert \BHyp \right)}
\newcommand{\BLikelihoodS}{\mathcal{L}\left( \BData \vert \BParm \right)}
\newcommand{\BLikelihoodL}{Pr \left( \BData \vert \BParm, \BHyp \right)}
\newcommand{\BPriorProbS}{\pi \left( \BParm \right)}
\newcommand{\BEvidenceL}{Pr \left( \BData \vert \BHyp \right)}
\newcommand{\BEvidenceS}{\mathcal{Z} \left( \BData \right)}

\newcommand{\RVpqAll}{V_{tpq\lambda}}
\newcommand{\RGpAll}{G_{tp\lambda}}
\newcommand{\RGqAll}{G_{tp\lambda}^H}
\newcommand{\REpAll}{E_{tps\lambda}}
\newcommand{\REqAll}{E_{tqs\lambda}^H}
\newcommand{\RKpAll}{K_{tps\lambda}}
\newcommand{\RKqAll}{K_{tqs\lambda}^H}
\newcommand{\RBAll}{B_{s\lambda}}

\begin{document}

\begin{frontmatter}

% Title Material
\title{Montblanc\tnoteref{t1}: GPU accelerated \\ Radio Interferometer Measurement Equations \\ in support of \\ Bayesian Inference for Radio Observations}
\tnotetext[t1]{\url{https://github.com/ska-sa/montblanc}}

\author[uctcs]{S.J.~Perkins\corref{cor1}}
%\author[uctcs]{S.J.~Perkins\corref{cor1}\fnref{fn1}}
\ead{sperkins@cs.uct.ac.za}
\author[uctcs]{P.C.~Marais}
\ead{patrick@cs.uct.ac.za}
\author[uwc,uctast]{Jonathan~T.~L.~Zwart}
\ead{jzwart@uwc.ac.za}
\author[uctast]{I. Natarajan}
\ead{iniyannatarajan@gmail.com}
\author[rhodesphy,gepi]{C.~Tasse}
\ead{cyril.tasse@obspm.fr}
\author[rhodes,skasa]{O.~Smirnov}
\ead{osmirnov@gmail.com}

\cortext[cor1]{Principal Corresponding Author}
%\fntext[fn1]{Author footnote text}
%\fntext[fn2]{Author footnote text}
%\fntext[fn3]{Author footnote text}
%\fntext[fn4]{Author footnote text}

\address[uctcs]{Department of Computer Science, University of Cape Town, Rondebosch, South Africa, 7700}
\address[rhodes]{Rhodes Centre for Radio Astronomy Techniques and Technologies, \\ Rhodes University, Grahamstown, South Africa, 6140}
\address[uwc]{Department of Physics \& Astronomy, University of the Western Cape, \\ Private Bag X17, Bellville, South Africa, 7535}
\address[uctast]{Department of Astronomy, University of Cape Town,  Rondebosch, South Africa, 7700}
\address[rhodesphy]{Department of Physics and Electronics, Rhodes University, PO Box 94,
Grahamstown, 6140 South Africa}
\address[gepi]{GEPI, Observatoire de Paris, CNRS, Universit\'e Paris Diderot, 5 place Jules Janssen, 92190 Meudon, France}
\address[skasa]{SKA South Africa, 3rd Floor, The Park, Park Road, Pinelands, 7405, South Africa}

\begin{keyword}
% See http://iopscience.iop.org/0004-637X/546/1/605/fulltext
% See http://www.acm.org/about/class
GPGPU -- computing methodologies: massively parallel algorithms -- Applied computing: Astronomy
 -- techniques:interferometric -- instrumentation: interferometers -- methods: statistical
\end{keyword}

\begin{abstract}
We present Montblanc, a GPU implementation
of the {\itshape Radio interferometer measurement equation} (RIME)
in support of the {\itshape Bayesian inference for radio observations} (BIRO)
technique. BIRO uses Bayesian inference to select sky models
that best match the visibilities observed by a radio interferometer.
To accomplish this, BIRO evaluates the RIME multiple times, varying
sky model parameters to produce multiple model visibilities. $\chi^2$
values computed from the model and observed visibilities
are used as likelihood values to drive the Bayesian sampling process and
select the best sky model.

As most of the elements of the RIME and $\chi^2$
calculation are independent of one another, they  are highly amenable
to parallel computation.
Additionally, Montblanc caters for
iterative RIME evaluation to produce multiple $\chi^2$ values.
Modified model parameters are transferred to the GPU between each iteration.

We implemented Montblanc as a Python package based upon
NVIDIA's CUDA architecture. As such, it is easy to extend and implement
different pipelines.
At present, Montblanc supports point and Gaussian morphologies, but is
designed for easy addition of new source profiles.

Montblanc's RIME implementation is performant:
On an NVIDIA K40, it is approximately 250 times faster
than MeqTrees on a dual hexacore Intel E5--2620v2 CPU.
Compared to the OSKAR simulator's GPU-implemented
RIME components it is 7.7 and 12 times faster on the same
K40 for single and double-precision floating point respectively.
However, OSKAR's RIME implementation is more general
than Montblanc's BIRO-tailored RIME.

Theoretical analysis of Montblanc's dominant CUDA kernel
suggests that it is memory bound. In practice, profiling shows
that is balanced between compute and memory,
as much of the data required by the problem
is retained in L1 and L2 cache.

\end{abstract}

\end{frontmatter}

%\maketitle

\section{Introduction}

\begin{figure}
\begin{center}
\subfloat[Antennas and baselines] {
    \includegraphics[width=0.31\columnwidth]{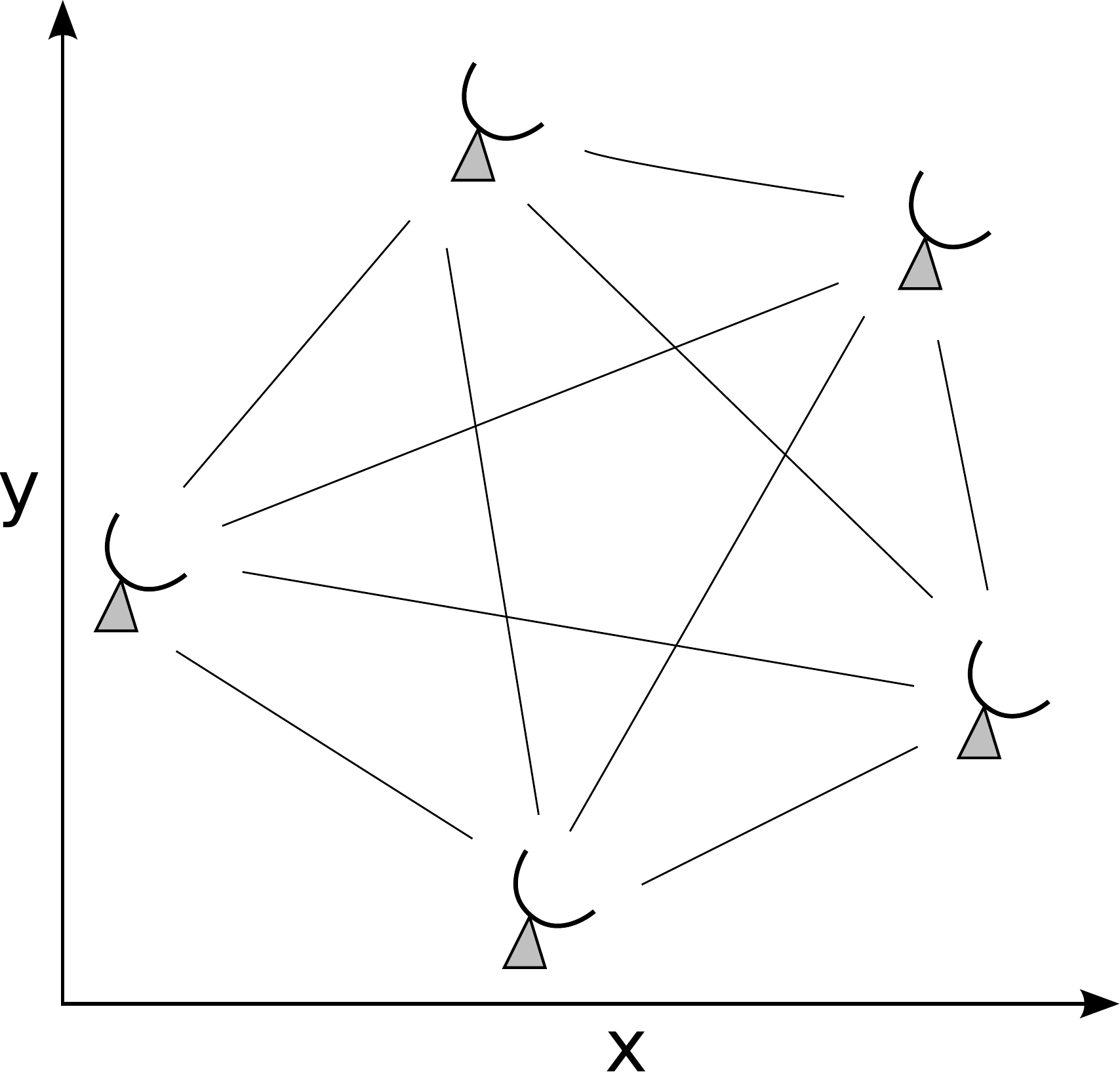}
    \label{fig:baselines}
}
\subfloat[Visibility tracks] {
    \includegraphics[width=0.31\columnwidth]{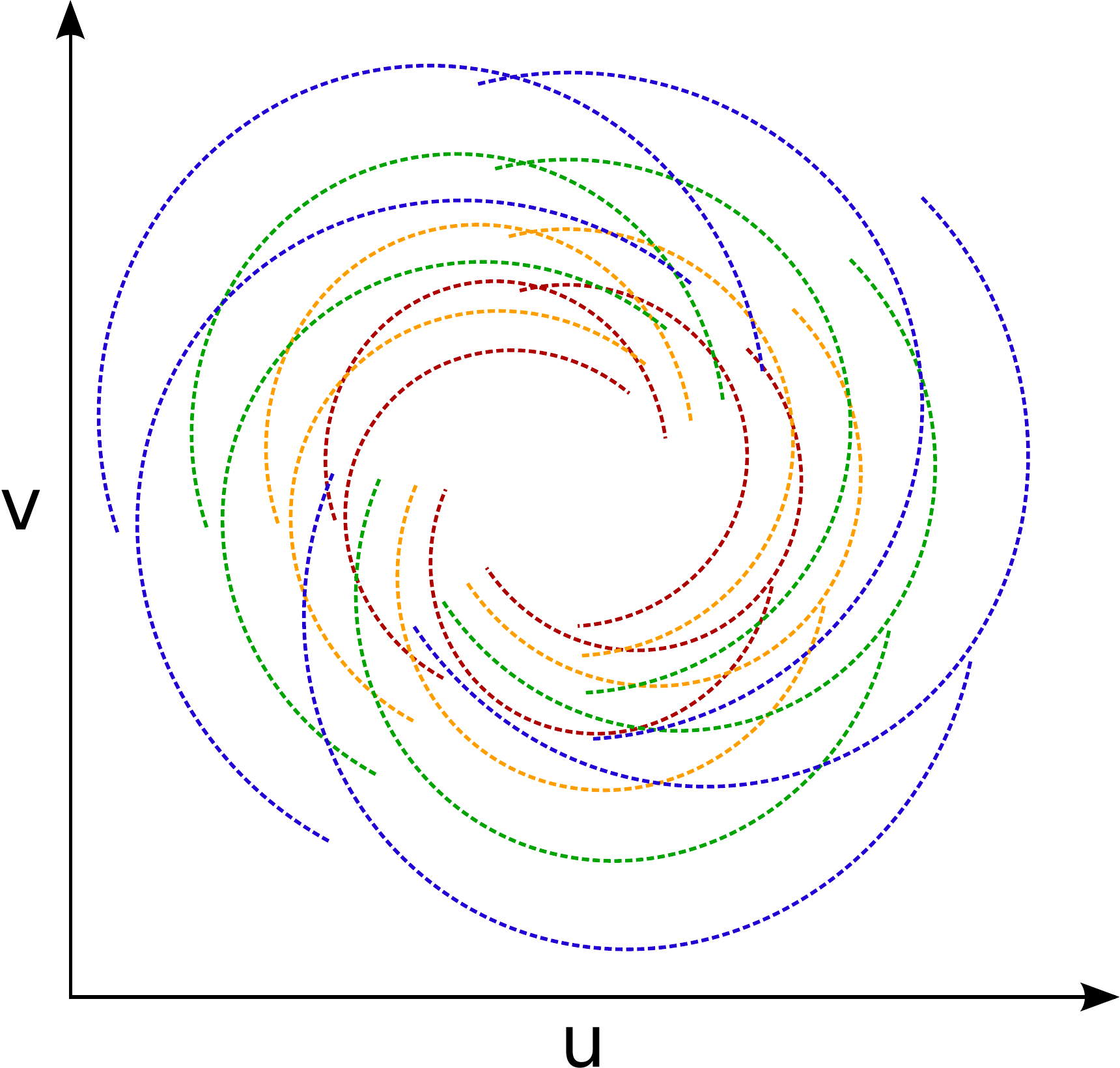}
    \label{fig:visibilities}
}
\subfloat[Sky model] {
    \includegraphics[width=0.31\columnwidth]{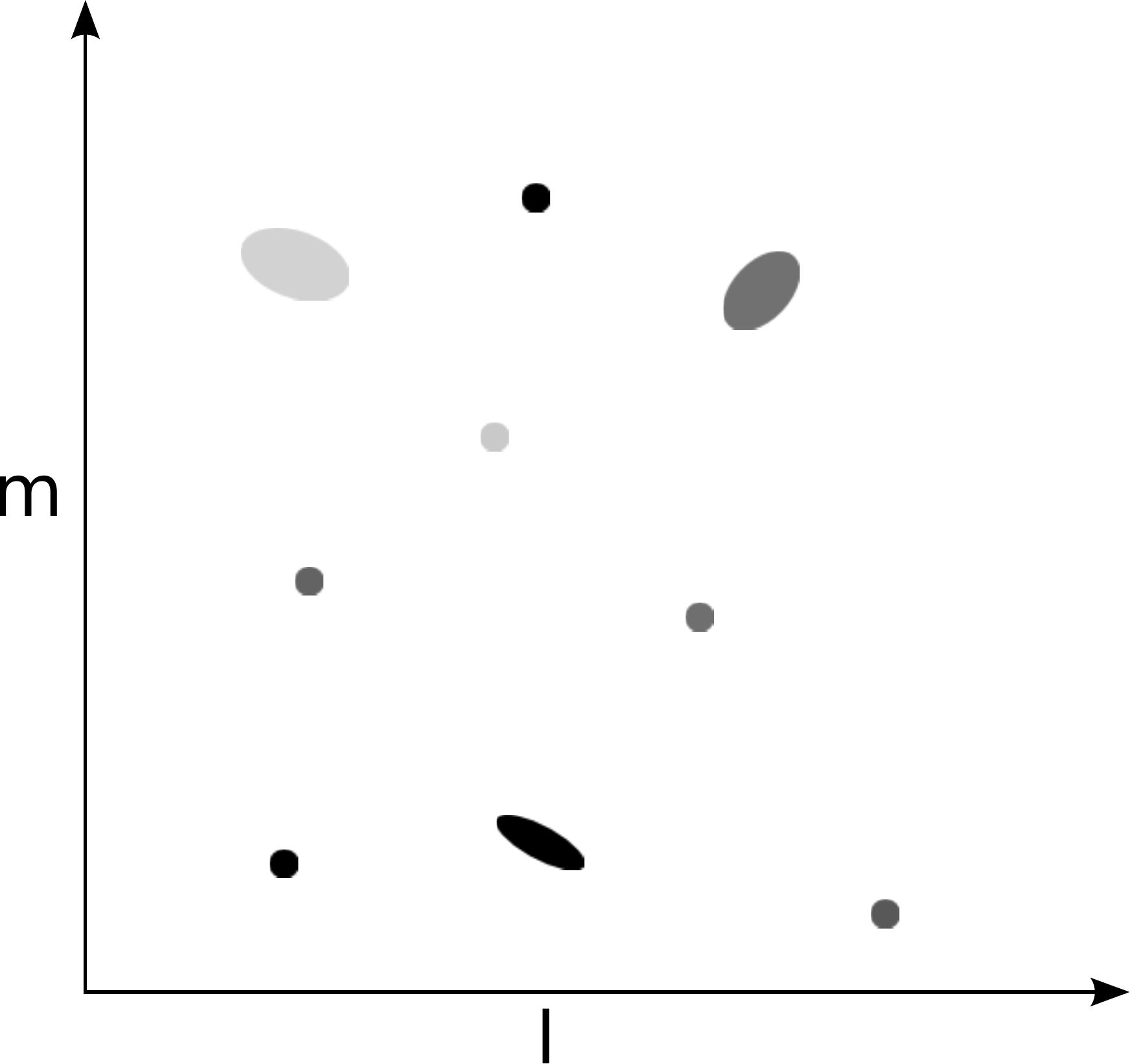}
    \label{fig:skymodel}
}
\end{center}
\caption{(a) Radio Interferometers are formed from multiple antennas.
Radio signals are measured and correlated on {\itshape baselines}
formed between antenna pairs.
(b) As the Earth rotates, {\itshape visibilities} are
observed in the spatial frequency domain, forming distinctive tracks.
(c) BIRO attempts to find a sky model, composed of point and
Gaussian sources, whose model visibilities are closest to
the interferometer's observed visibilities \citep{Thompson2007}.}
\end{figure}

\noindent {\itshape Bayesian inference for radio observations} (BIRO)
\citep{Lochner2015} is a rigorous statistical technique for inferring a
model of the sky brightness distribution from the observed
visibilities of a radio interferometer. Such models are composed of
various radio sources characterised by physical parameters. For
example, point sources are parameterised by position, flux density (in
total intensity and polarization) and a spectral index, while extended
sources require additional shape parameters. An example sky model
is shown in Figure \ref{fig:skymodel}.
BIRO explores the full posterior probability distribution
of the model parameters given the observed visibility data via
MCMC sampling.

Source extraction via imaging techniques must transform observed visibilities
(Figure \ref{fig:visibilities}) into a {\itshape dirty image}, that must be
processed by deconvolution algorithms such as \textsc{clean} \citep{Hogbom1974}.
Artifacts can be introduced by these algorithms when they attempt to
remove the synthesised beam.
The noise in this domain must be correlated and is not uniform across the
image. Sidelobes and choice of baseline weighting complicate analysis.
Sources are subsequently extracted from the map and characterized using,
for example, AIPS SAD, SExtractor\citep{Bertin1995} and PyBDSM\citep{Mohan2015}.
By contrast, BIRO uses Bayesian inference to find a sky model
that best fits the observed visibility data, given the measurement uncertainties.
Central to this process is the conversion of the sky model to model visibilities which,
in combination with observed visibilities, produce a $\chi^2$ value
(goodness of fit) that drives the Bayesian inference process.
Performing comparisons in the visibility domain is advantageous since
the noise is Gaussian, uncorrelated and stationary \citep{Feroz2009}.
BIRO therefore offers a simple, rigorous, powerful and flexible approach to
modelling the sky, bypassing the need for the deconvolution, imaging
and source-extraction steps.

The conversion of sky model to model visibilities is governed by
the Van Cittert–-Zernike theorem \citep{Thompson2007},
which provides the basis for modelling an interferometer's response
to a radio source.
Established models based on this theorem are only suited to
older calibration techniques that incorporate
direction-independent effects (DIEs).
However, they
are increasingly unsuitable for newer, more sensitive telescopes whose
wide-field/wide-band nature make them susceptible to subtler
direction-dependent effects (DDEs).
The Radio Interferometer Measurement Equation (RIME)
\citep{Hamaker1996,Smirnov2011_RIME1}
reformulates the Van Cittert--Zernike theorem using Jones calculus
and provides a rigorous basis for modelling current and future
interferometers and effects.
Due to this power and flexibility, it is
the RIME that BIRO makes use of to convert the
sky brightness distribution into model visibilities.
This even permits calibration to be undertaken
\textit{simultaneously} with modelling of the sky.

Existing BIRO implementations \cite{Lochner2015} use \textsc{MeqTrees}
\citep{Smirnov2010_MEQTREES} to evaluate the RIME, and (for example)
\textsc{MultiNEST} \citep{Feroz2009_MULTINEST} for performing Bayesian
parameter estimation and model selection.
However, the parameter space that BIRO must explore can be large
(10 to 100 parameters at present, 1000 to 10000 are planned in future)
 and often unusually shaped, resulting in tens of thousands of RIME
evaluations. Furthermore, a single RIME evaluation is expensive, since
visibilities must be calculated over time, baseline and channel, before
reduction to a single $\chi^2$ value.  Fortunately, these values can be
independently calculated, rendering the RIME particularly amenable to
a parallel implementation, accelerated by
Graphics Programming Units (GPUs).

The likelihood values used by {\itshape Bayesian inference }
are computed from the $\chi^2$ values and inform
the sky model selection process.
These calculations are not only relevant to BIRO.
Paraphrasing \citep{Kazemi2011}, telescope calibration
can be thought of as Maximum-likelihood (ML) estimation \cite{Mackay2003}
 of instrument and sky parameters through non-linear optimisation techniques,
such as Levenberg-Marquardt \citep{Levenberg1944, Marquardt1963}.
Radio astronomy packages such as CASA\citep{CASA2014}
and \textsc{MeqTrees} \citep{Smirnov2010_MEQTREES} implement ML
to perform calibration, and also rely on computation of
the RIME and $\chi^2$. Such techniques also
benefit from a fast RIME implementation.

OSKAR \citep{Mort2010} is a radio interferometer simulator
that implements certain parts of the RIME in NVIDIA's
CUDA \citep{CUDA} architecture and other parts in C. While
full-featured, it is a one-shot simulator and not designed for iterative
evaluation of the RIME, where only relevant sky parameters are changed
between BIRO iterations. Additionally, it does not currently support GPU
calculation of a $\chi^2$ value from the RIME and, due to the
specialist nature of the C programming language, it is not particularly
easy to add new source types, or extend.

This paper presents {\itshape Montblanc}, a GPU implementation of both
the RIME, and a calculator of the $\chi^2$ value from the RIME.
In contrast to OSKAR, we aim to implement the entire RIME on a GPU.
Implemented as a Python package, it in turn uses
the PyCUDA \citep{Kloeckner2012_PyCUDA} package
to access NVIDIA's CUDA architecture \citep{CUDA}.
Using Montblanc, the time required to calculate the RIME,
and the likelihood associated with a model parameter set is
dramatically reduced.
Montblanc currently supports both point and Gaussian sources, and is
architected for the addition of further source types, such as the
$\beta$-profile \citep{king66,king72}, the NFW
\citep{NFW} profile and corkscrews. A contributor \citep{Rivi2015}
has already added Sersic profiles \citep{sersic63} for instance.
 It also supports time-varying brightness and modelling of the beam profile,
 pointing solution and the noise covariance matrix. At present, it only supports the
expression of the DDE term as an analytic equation. DIE matrices are
not currently supported since BIRO generally operates
on fully calibrated data. Future support for these terms
would enable self-calibration simultaneously with sky model selection.

At present, Montblanc provides an {\itshape in-core}
solution to the RIME --- the problem size, generally governed
by the number of visibilities, must fit within the
RAM of a compute node. This is not currently problematic for BIRO
as time- and/or frequency-averaging (see e.g. \citep{Thompson2007})
can readily be applied post-flagging to compress the data
and hence reduce the computational load.
The degree of averaging that can be applied with loss of information
is set by the physics of the telescope in question
See e.g. \citep{Hobson2002, Feroz2009}.

The text is organised as follows: we first describe the RIME and
BIRO technique in more detail and provide an overview of GPU computing
and NVIDIA's CUDA architecture. This is followed by a discussion of
existing RIME implementations, \textsc{MeqTrees} and \textsc{OSKAR}.
We then describe Montblanc's architecture,
the process for computing the RIME, as well as the process
for subdividing the RIME so that the problem will fit within memory budgets.
We then present and discuss our results before concluding and mentioning
directions for future work.

\section{Background}
\label{sec:background}

\noindent
In this section we describe the Radio Interferometry Measurement Equation (RIME),
Bayesian Inference for Radio Observations (BIRO)
as well as GPUs and NVIDIA's CUDA architecture.

\subsection{The Radio Interferometry Measurement Equation}
\label{sec:rime}

\begin{table}
\begin{center}
\begin{tabular}{|l|l|l|l|}
\hline
name & number of & name & number of \\
\hline
ntime & timesteps & nsrc & sources \\
nbl & baselines & npsrc & point sources\\
na & antennas & ngsrc & gaussian sources\\
nchan & channels & & \\
\hline
\end{tabular}
\end{center}
\caption{RIME dimensions referred to in this work.}
\label{tbl:rime_dims}
\end{table}

\noindent
The Radio Interferometry Measurement Equation was developed by
Hamaker et.~al.~\citep{Hamaker1996} and revisited in a series of
papers by Smirnov \citep{Smirnov2011_RIME1}. It establishes a relation
between a sky brightness distribution and the response this
produces in an interferometer. The sky model is defined by radio
sources and the effects modifying their propagation along the line of
sight. By contrast, the interferometer response is measured as complex
voltages on an interferometer's correlated baselines.

The RIME is based on Jones calculus, originally developed to model the
polarisation of light by linear optical elements and applied here to
radio signals.
Jones calculus models radio signals as 2-element complex vectors
describing the transverse components of an EM plane wave, and
2x2 complex Jones matrices describing propagation effects.
Consequently, the RIME is well-suited to rigorously defining
signal propagation effects along the line of sight \citep{Smirnov2011_RIME2}.
We use the following formulation of the RIME over {\tt nsrc} sources:
\begin{align}
\RVpqAll = \RGpAll \left( \sum^{\textrm{nsrc}}_{s=0} \REpAll \RKpAll \RBAll \RKqAll \REqAll \right) \RGqAll,
\end{align}
where $\RVpqAll$ are the visibilities along the baseline formed by
antennas $p$ and $q$, at timestep $t$ and channel with wavelength
$\lambda$. $\RGpAll$ is antenna $p$'s DIE Jones matrix, $\REpAll$, the
DDE matrix for source $s$, $\RKpAll$, the phase matrix, and $\RBAll$,
the brightness matrix for source $s$. In general, $\RGpAll$ and
$\REpAll$ are matrices that may represent the linear product of more
specific propagation effects. A hermitian transpose ($H$) is applied to
the corresponding terms for antenna $q$. The formulation above expresses
the product and reduction of a 4D array of 2x2 matrices with
dimensions of time, antennas, baseline, sources and channel.
For the sake of readability we  sometimes exclude
time and channel in this paper.

While these terms are generally expressed as complex matrices and
 scalars, analytic expressions can be substituted, trading
memory storage, access and transfer costs for the computation of
a {\itshape tensor product}. For example, the brightness matrix
$\RBAll$ corresponding to astrophysical sources can typically
be parameterised as
\begin{align}
\RBAll = \left( \frac{\lambda_{ref}}{\lambda}\right)^\alpha
\left(
\begin{array}{cc}
I_s+Q_s & U_s+i V_s  \\
U_s-i V_s  & I_s-Q_s
\end{array}
\right), \label{eqn:brightness_term}
\end{align}
where $I_s,Q_s,U_s,V_s$ are Stokes parameters for source $s$ at
reference wavelength $\lambda_{ref}$.
Most extragalactic radio sources emit though the
synchrotron process, and their spectrum is modeled well by
Equation \ref{eqn:brightness_term} for moderately
wide frequency ranges.
To first order, most radio sources can be modelled extremely
comprehensively using a single spectral index $\alpha$.
This holds not only for the most common form of extragalactic emission
(synchroton radiation, $\alpha\simeq -0.7$), but also for thermal
(cosmic microwave background and/or Sunyaev--Zel'dovich effect)
and free-free radiation.
Higher-order spectral-index curvature terms, whose effects become
significant for moderately-wide bandwidths, could easily
be incorporated analytically.
\begin{align}
\RKpAll = \mathrm{e}^{\frac{2 \pi i}{\lambda}\left( u l + v m + w (n-1) \right)}, \label{eqn:k_term}
\end{align}
where  $u_{tp} = (u,v,w)$ is the \textit{uvw} coordinate for antenna $p$ at time $t$,
and $l_s = (l,m,n)$ the sky coordinate  for source $s$, with $n=\sqrt{1-l^2-m^2}$.
A common analytic approximation for
the primary beam profile $E_{ps}$ of the Westerbork Synthesis Radio Telescope
(WSRT)\citep{Popping2008,Smirnov2011_RIME2} is reasonably approximated as
\begin{align}
E_{ps} = \textrm{cos}^3\left(C\lambda \Vert l_s - \Delta l_p \Vert \right), \label{eqn:e_term}
\end{align}
where $\Delta l_p$ is the pointing error for antenna $p$ and
$C$=65\,GHz is a constant. While this expression applies to Westerbork,
others could be substituted, depending on the case.
For example, the VLA beam profile can be approximated with a
Jinc \citep{Smirnov2011_RIME2} function. Note that such
analytic expressions contain many expensive trigonometric and
transcendental functions.

\begin{figure}
\begin{center}
% Sketch output, version 0.3 (build 7d, Wed May 2 06:36:52 2012)
% Output language: PGF/TikZ,LaTeX
\begin{tikzpicture}[line join=round]
\filldraw[thin,fill=blue!20!white](-4.002,-3.286)--(-4.269,-3.393)--(-3.869,-3.464)--(-3.601,-3.357)--cycle;
\filldraw[thin,fill=blue!20!white](-4.002,.411)--(-4.269,.304)--(-3.869,.232)--(-3.601,.339)--cycle;
\filldraw[thin,fill=blue!20!white](-3.601,-3.357)--(-3.869,-3.464)--(-3.468,-3.536)--(-3.2,-3.429)--cycle;
\filldraw[thin,fill=blue!20!white](-4.269,-3.393)--(-4.537,-3.5)--(-4.136,-3.571)--(-3.869,-3.464)--cycle;
\filldraw[thin,fill=blue!20!white](-3.601,.339)--(-3.869,.232)--(-3.468,.161)--(-3.2,.268)--cycle;
\filldraw[thin,fill=blue!20!white](-3.2,-3.429)--(-3.468,-3.536)--(-3.067,-3.607)--(-2.8,-3.5)--cycle;
\filldraw[thin,fill=blue!20!white](-4.269,.304)--(-4.537,.196)--(-4.136,.125)--(-3.869,.232)--cycle;
\filldraw[thin,fill=blue!20!white](-3.869,-3.464)--(-4.136,-3.571)--(-3.735,-3.643)--(-3.468,-3.536)--cycle;
\filldraw[thin,fill=blue!20!white](-3.2,.268)--(-3.468,.161)--(-3.067,.089)--(-2.8,.196)--cycle;
\filldraw[thin,fill=blue!20!white](-4.537,-3.5)--(-4.804,-3.607)--(-4.403,-3.679)--(-4.136,-3.571)--cycle;
\filldraw[thin,fill=blue!20!white](-2.8,-3.5)--(-3.067,-3.607)--(-2.666,-3.679)--(-2.399,-3.571)--cycle;
\filldraw[thin,fill=blue!20!white](-3.869,.232)--(-4.136,.125)--(-3.735,.054)--(-3.468,.161)--cycle;
\filldraw[thin,fill=blue!20!white](-3.468,-3.536)--(-3.735,-3.643)--(-3.334,-3.714)--(-3.067,-3.607)--cycle;
\filldraw[thin,fill=blue!20!white](-2.8,.196)--(-3.067,.089)--(-2.666,.018)--(-2.399,.125)--cycle;
\filldraw[thin,fill=blue!20!white](-4.537,.196)--(-4.804,.089)--(-4.403,.018)--(-4.136,.125)--cycle;
\filldraw[thin,fill=blue!20!white](-1.998,-1.804)--(-1.998,-1.339)--(-2.265,-1.446)--(-2.265,-1.911)--cycle;
\filldraw[thin,fill=blue!20!white](-4.136,-3.571)--(-4.403,-3.679)--(-4.002,-3.75)--(-3.735,-3.643)--cycle;
\filldraw[thin,fill=blue!20!white](-2.399,-3.571)--(-2.666,-3.679)--(-2.265,-3.75)--(-1.998,-3.643)--cycle;
\filldraw[thin,fill=blue!20!white](-3.468,.161)--(-3.735,.054)--(-3.334,-.018)--(-3.067,.089)--cycle;
\filldraw[thin,fill=blue!20!white](-1.998,-1.339)--(-1.998,-.875)--(-2.265,-.982)--(-2.265,-1.446)--cycle;
\filldraw[thin,fill=blue!20!white](-4.804,-3.607)--(-5.071,-3.714)--(-4.67,-3.786)--(-4.403,-3.679)--cycle;
\filldraw[thin,fill=blue!20!white](-1.998,-4.107)--(-1.998,-3.643)--(-2.265,-3.75)--(-2.265,-4.214)--cycle;
\filldraw[thin,fill=blue!20!white](-3.067,-3.607)--(-3.334,-3.714)--(-2.933,-3.786)--(-2.666,-3.679)--cycle;
\filldraw[thin,fill=blue!20!white](-4.136,.125)--(-4.403,.018)--(-4.002,-.054)--(-3.735,.054)--cycle;
\filldraw[thin,fill=blue!20!white](-2.399,.125)--(-2.666,.018)--(-2.265,-.054)--(-1.998,.054)--cycle;
\filldraw[thin,fill=blue!20!white](-1.998,-.875)--(-1.998,-.411)--(-2.265,-.518)--(-2.265,-.982)--cycle;
\filldraw[thin,fill=blue!20!white](-3.735,-3.643)--(-4.002,-3.75)--(-3.601,-3.821)--(-3.334,-3.714)--cycle;
\filldraw[thin,fill=blue!20!white](-3.067,.089)--(-3.334,-.018)--(-2.933,-.089)--(-2.666,.018)--cycle;
\filldraw[thin,fill=blue!20!white](-1.998,-.411)--(-1.998,.054)--(-2.265,-.054)--(-2.265,-.518)--cycle;
\filldraw[thin,fill=blue!20!white](-2.265,-1.911)--(-2.265,-1.446)--(-2.532,-1.554)--(-2.532,-2.018)--cycle;
\filldraw[thin,fill=blue!20!white](-4.804,.089)--(-5.071,-.018)--(-4.67,-.089)--(-4.403,.018)--cycle;
\filldraw[thin,fill=blue!20!white](-4.403,-3.679)--(-4.67,-3.786)--(-4.269,-3.857)--(-4.002,-3.75)--cycle;
\filldraw[thin,fill=blue!20!white](-2.666,-3.679)--(-2.933,-3.786)--(-2.532,-3.857)--(-2.265,-3.75)--cycle;
\filldraw[thin,fill=blue!20!white](.33,-1.946)--(.33,-1.482)--(.062,-1.589)--(.062,-2.054)--cycle;
\filldraw[thin,fill=blue!20!white](-3.735,.054)--(-4.002,-.054)--(-3.601,-.125)--(-3.334,-.018)--cycle;
\filldraw[thin,fill=blue!20!white](-2.265,-1.446)--(-2.265,-.982)--(-2.532,-1.089)--(-2.532,-1.554)--cycle;
\filldraw[thin,fill=blue!20!white](-2.265,-4.214)--(-2.265,-3.75)--(-2.532,-3.857)--(-2.532,-4.321)--cycle;
\filldraw[thin,fill=blue!20!white](-3.334,-3.714)--(-3.601,-3.821)--(-3.2,-3.893)--(-2.933,-3.786)--cycle;
\filldraw[thin,fill=blue!20!white](-5.071,-3.714)--(-5.339,-3.821)--(-4.938,-3.893)--(-4.67,-3.786)--cycle;
\filldraw[thin,fill=blue!20!white](-2.666,.018)--(-2.933,-.089)--(-2.532,-.161)--(-2.265,-.054)--cycle;
\filldraw[thin,fill=blue!20!white](-4.403,.018)--(-4.67,-.089)--(-4.269,-.161)--(-4.002,-.054)--cycle;
\filldraw[thin,fill=blue!20!white](-5.339,-1.982)--(-4.938,-2.054)--(-4.938,-1.589)--(-5.339,-1.518)--cycle;
\filldraw[thin,fill=blue!20!white](-2.265,-.982)--(-2.265,-.518)--(-2.532,-.625)--(-2.532,-1.089)--cycle;
\filldraw[thin,fill=blue!20!white](.33,-1.482)--(.33,-1.018)--(.062,-1.125)--(.062,-1.589)--cycle;
\filldraw[thin,fill=blue!20!white](-.339,-1.982)--(.062,-2.054)--(.062,-1.589)--(-.339,-1.518)--cycle;
\filldraw[thin,fill=blue!20!white](-4.002,-3.75)--(-4.269,-3.857)--(-3.869,-3.929)--(-3.601,-3.821)--cycle;
\filldraw[thin,fill=blue!20!white](.33,-1.018)--(.33,-.554)--(.062,-.661)--(.062,-1.125)--cycle;
\filldraw[thin,fill=blue!20!white](-.071,-.018)--(-.339,-.125)--(.062,-.196)--(.33,-.089)--cycle;
\filldraw[thin,fill=blue!20!white](-.339,-1.518)--(.062,-1.589)--(.062,-1.125)--(-.339,-1.054)--cycle;
\filldraw[thin,fill=blue!20!white](-3.334,-.018)--(-3.601,-.125)--(-3.2,-.196)--(-2.933,-.089)--cycle;
\filldraw[thin,fill=blue!20!white](-5.071,-.018)--(-5.339,-.125)--(-4.938,-.196)--(-4.67,-.089)--cycle;
\filldraw[thin,fill=blue!20!white](-2.532,-2.018)--(-2.532,-1.554)--(-2.8,-1.661)--(-2.8,-2.125)--cycle;
\filldraw[thin,fill=blue!20!white](-5.339,-1.518)--(-4.938,-1.589)--(-4.938,-1.125)--(-5.339,-1.054)--cycle;
\filldraw[thin,fill=blue!20!white](-2.265,-.518)--(-2.265,-.054)--(-2.532,-.161)--(-2.532,-.625)--cycle;
\filldraw[thin,fill=blue!20!white](-2.933,-3.786)--(-3.2,-3.893)--(-2.8,-3.964)--(-2.532,-3.857)--cycle;
\filldraw[thin,fill=blue!20!white](-4.67,-3.786)--(-4.938,-3.893)--(-4.537,-3.964)--(-4.269,-3.857)--cycle;
\filldraw[thin,fill=blue!20!white](-5.339,-4.286)--(-4.938,-4.357)--(-4.938,-3.893)--(-5.339,-3.821)--cycle;
\filldraw[thin,fill=blue!20!white](-4.938,-2.054)--(-4.537,-2.125)--(-4.537,-1.661)--(-4.938,-1.589)--cycle;
\filldraw[thin,fill=blue!20!white](-5.339,-1.054)--(-4.938,-1.125)--(-4.938,-.661)--(-5.339,-.589)--cycle;
\filldraw[thin,fill=blue!20!white](.33,-.554)--(.33,-.089)--(.062,-.196)--(.062,-.661)--cycle;
\filldraw[thin,fill=blue!20!white](-.339,-1.054)--(.062,-1.125)--(.062,-.661)--(-.339,-.589)--cycle;
\filldraw[thin,fill=blue!20!white](-4.002,-.054)--(-4.269,-.161)--(-3.869,-.232)--(-3.601,-.125)--cycle;
\filldraw[thin,fill=blue!20!white](-2.532,-1.554)--(-2.532,-1.089)--(-2.8,-1.196)--(-2.8,-1.661)--cycle;
\filldraw[thin,fill=blue!20!white](-2.532,-4.321)--(-2.532,-3.857)--(-2.8,-3.964)--(-2.8,-4.429)--cycle;
\filldraw[thin,fill=blue!20!white](-3.601,-3.821)--(-3.869,-3.929)--(-3.468,-4)--(-3.2,-3.893)--cycle;
\filldraw[thin,fill=blue!20!white](-2.532,-1.089)--(-2.532,-.625)--(-2.8,-.732)--(-2.8,-1.196)--cycle;
\filldraw[thin,fill=blue!20!white](-4.938,-1.589)--(-4.537,-1.661)--(-4.537,-1.196)--(-4.938,-1.125)--cycle;
\filldraw[thin,fill=blue!20!white](-4.67,-.089)--(-4.938,-.196)--(-4.537,-.268)--(-4.269,-.161)--cycle;
\filldraw[thin,fill=blue!20!white](-2.933,-.089)--(-3.2,-.196)--(-2.8,-.268)--(-2.532,-.161)--cycle;
\filldraw[thin,fill=blue!20!white](-.339,-.589)--(.062,-.661)--(.062,-.196)--(-.339,-.125)--cycle;
\filldraw[thin,fill=blue!20!white](-5.339,-.589)--(-4.938,-.661)--(-4.938,-.196)--(-5.339,-.125)--cycle;
\filldraw[thin,fill=blue!20!white](-4.938,-4.357)--(-4.537,-4.429)--(-4.537,-3.964)--(-4.938,-3.893)--cycle;
\filldraw[thin,fill=blue!20!white](-4.269,-3.857)--(-4.537,-3.964)--(-4.136,-4.036)--(-3.869,-3.929)--cycle;
\filldraw[thin,fill=blue!20!white](-2.8,-2.125)--(-2.8,-1.661)--(-3.067,-1.768)--(-3.067,-2.232)--cycle;
\filldraw[thin,fill=blue!20!white](-4.938,-1.125)--(-4.537,-1.196)--(-4.537,-.732)--(-4.938,-.661)--cycle;
\filldraw[thin,fill=blue!20!white](-2.532,-.625)--(-2.532,-.161)--(-2.8,-.268)--(-2.8,-.732)--cycle;
\filldraw[thin,fill=blue!20!white](-4.537,-2.125)--(-4.136,-2.196)--(-4.136,-1.732)--(-4.537,-1.661)--cycle;
\filldraw[thin,fill=blue!20!white](-3.601,-.125)--(-3.869,-.232)--(-3.468,-.304)--(-3.2,-.196)--cycle;
\filldraw[thin,fill=blue!20!white](-3.2,-3.893)--(-3.468,-4)--(-3.067,-4.071)--(-2.8,-3.964)--cycle;
\filldraw[thin,fill=blue!20!white](-4.938,-.661)--(-4.537,-.732)--(-4.537,-.268)--(-4.938,-.196)--cycle;
\filldraw[thin,fill=blue!20!white](-4.269,-.161)--(-4.537,-.268)--(-4.136,-.339)--(-3.869,-.232)--cycle;
\filldraw[thin,fill=blue!20!white](-4.537,-1.661)--(-4.136,-1.732)--(-4.136,-1.268)--(-4.537,-1.196)--cycle;
\filldraw[thin,fill=blue!20!white](-2.8,-1.661)--(-2.8,-1.196)--(-3.067,-1.304)--(-3.067,-1.768)--cycle;
\filldraw[thin,fill=blue!20!white](-4.537,-4.429)--(-4.136,-4.5)--(-4.136,-4.036)--(-4.537,-3.964)--cycle;
\filldraw[thin,fill=blue!20!white](-2.8,-4.429)--(-2.8,-3.964)--(-3.067,-4.071)--(-3.067,-4.536)--cycle;
\filldraw[thin,fill=blue!20!white](-3.869,-3.929)--(-4.136,-4.036)--(-3.735,-4.107)--(-3.468,-4)--cycle;
\filldraw[thin,fill=blue!20!white](-4.136,-2.196)--(-3.735,-2.268)--(-3.735,-1.804)--(-4.136,-1.732)--cycle;
\filldraw[thin,fill=blue!20!white](-3.2,-.196)--(-3.468,-.304)--(-3.067,-.375)--(-2.8,-.268)--cycle;
\filldraw[thin,fill=blue!20!white](-2.8,-1.196)--(-2.8,-.732)--(-3.067,-.839)--(-3.067,-1.304)--cycle;
\filldraw[thin,fill=blue!20!white](-4.537,-1.196)--(-4.136,-1.268)--(-4.136,-.804)--(-4.537,-.732)--cycle;
\filldraw[thin,fill=blue!20!white](-2.8,-.732)--(-2.8,-.268)--(-3.067,-.375)--(-3.067,-.839)--cycle;
\filldraw[thin,fill=blue!20!white](-3.067,-2.232)--(-3.067,-1.768)--(-3.334,-1.875)--(-3.334,-2.339)--cycle;
\filldraw[thin,fill=blue!20!white](-3.869,-.232)--(-4.136,-.339)--(-3.735,-.411)--(-3.468,-.304)--cycle;
\filldraw[thin,fill=blue!20!white](-4.136,-1.732)--(-3.735,-1.804)--(-3.735,-1.339)--(-4.136,-1.268)--cycle;
\filldraw[thin,fill=blue!20!white](-4.537,-.732)--(-4.136,-.804)--(-4.136,-.339)--(-4.537,-.268)--cycle;
\filldraw[thin,fill=blue!20!white](-4.136,-4.5)--(-3.735,-4.571)--(-3.735,-4.107)--(-4.136,-4.036)--cycle;
\filldraw[thin,fill=blue!20!white](-3.468,-4)--(-3.735,-4.107)--(-3.334,-4.179)--(-3.067,-4.071)--cycle;
\filldraw[thin,fill=blue!20!white](-3.735,-2.268)--(-3.334,-2.339)--(-3.334,-1.875)--(-3.735,-1.804)--cycle;
\filldraw[thin,fill=blue!20!white](-3.067,-1.768)--(-3.067,-1.304)--(-3.334,-1.411)--(-3.334,-1.875)--cycle;
\filldraw[thin,fill=blue!20!white](-4.136,-1.268)--(-3.735,-1.339)--(-3.735,-.875)--(-4.136,-.804)--cycle;
\filldraw[thin,fill=blue!20!white](-3.067,-4.536)--(-3.067,-4.071)--(-3.334,-4.179)--(-3.334,-4.643)--cycle;
\filldraw[thin,fill=blue!20!white](-3.067,-1.304)--(-3.067,-.839)--(-3.334,-.946)--(-3.334,-1.411)--cycle;
\filldraw[thin,fill=blue!20!white](-3.735,-1.804)--(-3.334,-1.875)--(-3.334,-1.411)--(-3.735,-1.339)--cycle;
\filldraw[thin,fill=blue!20!white](-3.468,-.304)--(-3.735,-.411)--(-3.334,-.482)--(-3.067,-.375)--cycle;
\filldraw[thin,fill=blue!20!white](-4.136,-.804)--(-3.735,-.875)--(-3.735,-.411)--(-4.136,-.339)--cycle;
\filldraw[thin,fill=blue!20!white](-3.735,-4.571)--(-3.334,-4.643)--(-3.334,-4.179)--(-3.735,-4.107)--cycle;
\filldraw[thin,fill=blue!20!white](-3.735,-1.339)--(-3.334,-1.411)--(-3.334,-.946)--(-3.735,-.875)--cycle;
\filldraw[thin,fill=blue!20!white](-3.067,-.839)--(-3.067,-.375)--(-3.334,-.482)--(-3.334,-.946)--cycle;
\filldraw[thin,fill=blue!20!white](-3.735,-.875)--(-3.334,-.946)--(-3.334,-.482)--(-3.735,-.411)--cycle;
\path[-latex] (-5.138,-4.6) -- (-4.336,-4.743) node[sloped,pos=0.6,below] {$\mathbf{na}$};\path[-latex] (-3.134,-5.236) -- (-2.599,-5.021) node[sloped,pos=1.1,above] {$\mathbf{ntime}$};\path[-latex] (-.539,-1.714) -- (-.539,-.971) node[sloped,pos=0.6,above] {$\mathbf{nsrc}$};\node at (-3,-6) {$u_{tp}$};\node at (1,-1) {$l_{s}$};\node at (-6,.5) {$\RKpAll$};\end{tikzpicture}% End sketch output
\end{center}
\caption{The tensor product, $\RKpAll$, is created by
combining the uvw coordinate $u_{tp}$,
the source coordinate $l_{s}$
and the wavelength $\lambda$ (not shown).}
\label{fig:tensor_product}
\end{figure}
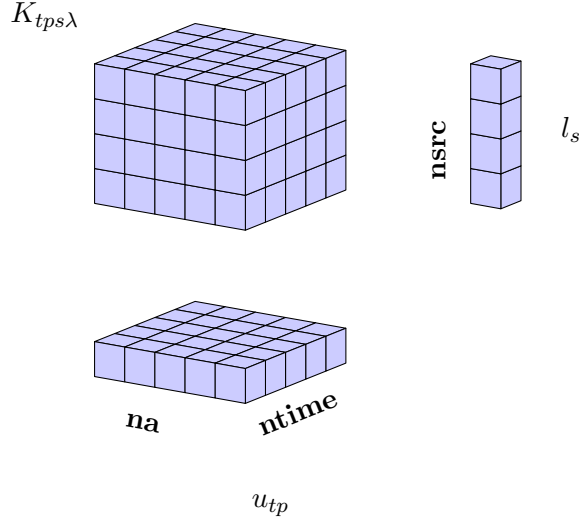

Finally, the {\itshape source coherency} is defined as
\begin{align}
X_{pqs} = E_{ps} K_{ps} B_s K_{qs}^H E_{qs}^H,
\end{align}
and by summing over the number of sources, we can
obtain the {\itshape model visibilities} along baseline $pq$:
\begin{align}
X_{pq} = \sum^{nsrc}_{s=0} X_{pqs}.
\end{align}
The model visibilities produced by the RIME can then
be compared against the actual observed visibilities
of the telescope (see below).

\subsection{RIME Dimensions and Parallelism}

\noindent
It is useful to consider the dimensionality of the RIME, since this
reveals the inherent parallelism of the equation. As discussed in
the previous section, the primary dimensions involved
in solving the RIME are timesteps, antennas, baselines, sources and
channels. We refer to these dimensions as {\tt ntime}, {\tt na}, {\tt
  nbl}, {\tt nsrc} and {\tt nchan} respectively (Table \ref{tbl:rime_dims}).

Consider the phase term, $\RKpAll = \exp\left[(2
\pi/\lambda)(u_{tp} \cdot l_s)\right]$ with input wavelength $\lambda$
($\mathtt{nchan}$), \textit{uvw} coordinate $u_{tp}$ ($\mathtt{ntime
  \times na}$) and source coordinate $l_s$ ($\mathtt{nsrc}$). Then,
the tensor product of these inputs $\RKpAll$ (Figure \ref{fig:tensor_product}),
has dimension $\mathtt{ntime \times na \times nsrc \times nchan}$. The values of
this 4D tensor can be computed entirely independently of one another.

Per-antenna terms $\RKpAll$ and $\RKqAll$ with dimension $\mathtt{ntime
  \times na \times nsrc \times nchan}$ are combined to produce a
$\mathtt{ntime \times nbl \times nsrc \times nchan}$ per-baseline term
$K_{tpqs\lambda}$: Each antenna is combined with every other antenna to form
$\mathtt{nbl} = \mathtt{na}(\mathtt{na}-1)/2$ baselines, discarding
auto-correlations where $p=q$.

Thus to solve the RIME, a $\mathtt{ntime \times nbl \times nsrc \times
  nchan}$ source-coherency array is computed from the product of tensors.
The source dimension of this array is reduced
to produce a \linebreak[4] $\mathtt{ntime \times nbl \times nchan}$
model visibility array. Each value of the source coherency, and each
sum over its three dimensions can be computed independently. It is the
independence inherent in this calculation, and the dimensionality of the
problem, that make the RIME particularly suitable to parallel
implementation (section \ref{sec:implementation}).

\subsection{Bayesian Inference for Radio Observations}

\noindent
{\itshape Bayesian inference} is a method for estimating the set of
model parameters $\BParm$ describing a model or hypothesis $\BHyp$
that includes any assumptions, for data $\BData$. Bayes' theorem states that
\begin{align}
\BPosteriorProbL = \frac{\BLikelihoodL \BPriorProbL}{\BEvidenceL}.
\end{align}
$\BPosteriorProbL \equiv \BPosteriorProbS$ is the {\itshape posterior}
probability distribution of the parameters, $\BLikelihoodL \equiv
\BLikelihoodS$ the {\itshape likelihood}, $\BPriorProbL \equiv
\BPriorProbS$ the {\itshape prior} probability distribution and
$\BEvidenceL \equiv \BEvidenceS$ the {\itshape evidence}.

The posterior $\BPosteriorProbS$ is the probability distribution of
the parameters for the hypothesis or model. In the case of BIRO
this refers to the probability that
the parameter set governing a particular configuration/instance of
the RIME produces model visibilities that match the observed
visibilities. The likelihood $\BLikelihoodS$ is the probability that
the data or model is correct, given a set of parameters. Under the
assumption of Gaussian measurement uncertainties, this is
a $\chi^2$ value created by comparing the model visibilities with the
observed visibilities $D_{pq}$, given a weight vector $w_{pq}$:
\begin{align}
\label{eqn:lhood}
-2\ln \BLikelihoodS = \sum w_{pq} \left(V_{pq} - D_{pq}\right)^2
+\sum \ln \left( 2\pi/w_{pq}\right)=\chi^2 +\sum \ln \left( 2\pi/w_{pq}\right).
\end{align}
The prior $\BPriorProbS$ is the pre-existing probability that
the parameter set matches the hypothesis or model, while,
$\BEvidenceS$ described below, is the evidence for the data.

Similar to the RIME, this process of calculating the $\chi^2$ value
involves differencing, squaring and multiplying the individual
elements of three $\mathtt{ntime  \times nbl \times nchan}$) arrays.
As such, it is also a highly parallel operation.

Bayesian inference can also be used to perform model selection. In
BIRO's case for example, this involves choosing the model of the sky that best
matches the observed visibilities. This may be a particular
combination of point and Gaussian sources. Here the objective is to
find the evidence $\BEvidenceS$, the factor required to normalise the
posterior over parameter space $\BParm$:
\begin{align}
\BEvidenceS = \int \BLikelihoodS \BPriorProbS d^D \BParm,
\end{align}
where $D$ is the dimensionality of parameter space. A salient feature
of Bayesian model selection is that it follows Occam's Razor: it chooses
more compact parameter spaces over larger ones, rewarding a good fit
but penalizing more complex models. When comparing two models, the
ratio of the posteriors $R$, given the observed data set $\BData$, is used
to select the one with the higher probability:
\begin{align}
R = \frac{Pr(H_1 \vert \BData)}{Pr(H_2 \vert \BData)}.
\end{align}

\subsection{GPUs and CUDA}
\label{sec:gpu_cuda}

\noindent
Prior to 2005 \citep{Barsdell2010}, computing power followed
Moore's Law \citep{Moore1965}, increasing in direct proportion to
increases in Central Processing Unit (CPU) clock rates.
However, physical limitations
in the manufacturing process prevent further advances
with this strategy. Chip manufacturers
therefore resorted to providing further compute power
by introducing multiple cores on a CPU. Dual, quad, hexacore and
octocore CPUs are now ubiquitous.

This strategy is most evident in the architecture of
Graphics Programming Units (GPU).  GPUs have thousands of cores
devoted to rendering billions of pixels. Tens of thousands of threads
execute a common shader program on separate pixel data.
These shaders are classed as
Single Instruction, Multiple Data Streams (SIMD) in
Flynn's Taxonomy \citep{Flynn1972} and their computation
is highly parallel.

NVIDIA's Compute Unified Device Architecture (CUDA) \citep{CUDA}
generalises GPU programming to non-graphics related
programs.The parallel nature of many Radio Astronomy
\citep{Barsdell2010} algorithms make them amenable to
GPU implementation.
Instead of shader programs, CUDA {\itshape kernels} are
written in a variant of C. Each kernel executes many threads,
grouped into separate {\itshape thread blocks} for execution
on GPU cores. Each core executes a group of 32 threads
called a {\itshape warp}.

The NVIDIA Kepler K40 consists of
15 {\itshape streaming multiprocessors} (SMX), each
containing 192 cores for a total of 2880 cores. Each SMX
has 65,536 registers, 64KB of L1 memory, split
16KB/48KB between L1 cache and shared memory
configurations, as well as a 48KB read-only L1 cache.
An additional 1536KB of L2 cache is shared
by all SMXs \citep{KeplerWhitePaper}.

The amount of theoretical processing power of
GPUs is high: A K40's peak
single floating point performance is 4290 GFLOPS/s,
more than $10\times$ the 371 GFLOPS/s
available to a dual 2.90GHz Intel E5--2690 CPU.
In practice, such performance can only be attained by
kernels composed of {\itshape fused multiply-adds} (FMA),
performed on data in {\itshape registers} and
most kernels will not conform to this pattern.
Additionally, processor speeds have increased at a far greater
rate, compared to memory access speeds. Many operations
can occur while waiting for a memory read, but this in turn leads
to situations where data dependent algorithms
``starve'' a processor  \citep{Alted2010}.
To ameliorate this, GPUs support high memory bandwidth:
memory read latency is amortised via large data transfers
and processors waiting for data are assigned other computation.
A K40 can read 288 GB/s and the NVIDIA Pascal architecture
is projected to provide $\approx 12000$ GFLOPS/s with
a memory bandwidth of $\approx 1000$ GB/s.
By contrast a E5--2690 has a bandwidth of 51.2 GB/s.

Therefore, approaching a GPU's (or CPU's) peak performance is
highly dependent on an algorithm's {\itshape arithmetic intensity}
-- the number of FLOPS performed per bytes read. To approach
this performance, it is necessary to exploit a device's
memory architecture as far as possible.
In practice, this means  ameliorating and avoiding high latency reads
of the GPU's global memory, firstly by ensuring that threads access contiguous
memory locations to achieve {\itshape coalesced reads}. Secondly,
as much of the problem as possible should be retained in
fast {\itshape registers}, {\itshape shared memory} and {\itshape cache}.

Due to the limited registers and shared memory per SMX,
 there is a trade-off between the number of threads executed
 on a SMX, and the registers and shared memory used by each thread.
It is desirable for a kernel to exhibit high {\itshape occupancy} by
executing as many threads as possible on a SMX since this fully
exploits a device's parallelism. However, high occupancy tends
to limit (performant) kernel complexity, as the amount of
registers and shared memory available to each thread is reduced.

\section{Previous Work}

\noindent
Both \textsc{MeqTrees} \citep{Smirnov2010_MEQTREES}
and \textsc{OSKAR} \citep{Mort2010} are packages
that implement the RIME in order to generate visibilities
from a parametric sky model.

The goal of \textsc{MeqTrees} is to provide a tool for the
rapid development of Radio Astronomy models.
It is primarily designed to evaluate the RIME for purposes
of telescope simulation and/or calibration the RIME,
through the use of {\itshape expression trees}.
This representation supports
decomposition of the RIME into separate pieces of work
for parallel execution over multiple CPU threads.
Its back-end is implemented in C++, while Python allows for
developers to rapidly prototype on the front-end.

\textsc{OSKAR} is implemented in both C and CUDA.
\textsc{OSKAR}'s solution to
the RIME is designed for generality -- some RIME
components are implemented as separate CUDA kernels,
while others are implemented on the CPU. The results from
these components are multiplied together by CUDA kernels
or CPU functions and correlated with one another.

In contrast to direct evaluation of the RIME
(which is effectively a DFT), it is possible to use
an FFT in combination with convolutional degridding
\citep{Cornwell2008} to obtain model visibilities,
given a rasterised sky model image.
For example, CASA's \citep{CASA2014} imaging
functionality could be used to calculate a $\chi^2$,
although this is not currently implemented in the API.

Both the MeqTrees and OSKAR2 simulators are mature, but are
not designed to iteratively evaluate the RIME
on a GPU. This is not ideal
for BIRO's case, since, in order to calculate
the many $\chi^2$ values associated with
different parameters, multiple, iterative
RIME evaluations are necessary each
producing model visibilities.
Additionally, they do not, as yet, have the facility
to compute a $\chi^2$ from model and
observed visibilities at each step of the sampling chain.
This highly parallel computation would also benefit from
GPU implementation.

The FFT approach is not ideal for BIRO
for a several reasons.
First, discretising/rasterising radio sources defined
by continuous parameters requires rasterisation and this
image must be continually recreated to account
for changing source parameters.
At present, the effect of source averaging
on the parametric process is not well-understood.
Second, degridding must be applied to extract
per-time, -baseline and -channel visibilities from
the gridded visibilities, introducing further
averaging error into the process.
Third, while the degridding approach has an inexpensive FFT
--- $O(N^2 \, \textrm{log} \,  N^2)$ for an $N \times N$ image ---
the computational complexity of degridding itself is
$O(\mathtt{nvis \times c^2})$ vs
$O(\mathtt{nvis \times nsrc})$ for the RIME
(see Section \ref{sec:kernel_times_comp_complexity})
where {\tt nvis}, {\tt c} and {\tt nsrc}
are the number of visibilities, convolution support size and
number of sources, respectively.
Thus, the complexity of the two approaches
only differ by the square of the
convolution support size and
the number of sky model sources ---
a $256 \times 256$ kernel corresponds
to 65,536 sources, for instance. This
number of sources is more than reasonable
for current BIRO requirements, as each of their
parameters must be allowed to vary.
The degridding approach may indeed be useful for
scenarios where radio sources can not be
described analytically, or for cases
involving many faint sources.
However, BIRO does not at present support
unparameterised radio-sources.

\begin{table}
\begin{center}
\begin{tabular}{|l|l|l|l|}
\hline
name & dimensions & type & from MS \\
\hline
{\itshape uvw} & $\mathtt{ntime \times na}$ & float & $\checkmark$ \\
antenna pairs & $\mathtt{ntime \times nbl}$ & integer & $\checkmark$\\
{\itshape lm} & $\mathtt{nsrc}$ & float & $\times$ \\
brightness & $\mathtt{ntime \times nsrc}$ & float & $\times$ \\
gaussian shapes & $\mathtt{ngsrc}$ & float & $\times$ \\
wavelength & $\mathtt{nchan}$ & float & $\checkmark$ \\
pointing errors & $\mathtt{ntime \times na}$ & float & $\times$ \\
weight vector & $\mathtt{ntime \times nbl \times nchan}$ & float & $\checkmark$ \\
data & $\mathtt{ntime \times nbl \times nchan}$ & complex & $\checkmark$ \\
\hline
\end{tabular}
\end{center}
\caption{Input to the RIME. Depending on the required accuracy,
float or complex types may have single or double floating point precision.}
\label{tbl:rime_input}
\end{table}

\begin{figure}
\begin{center}
\includegraphics{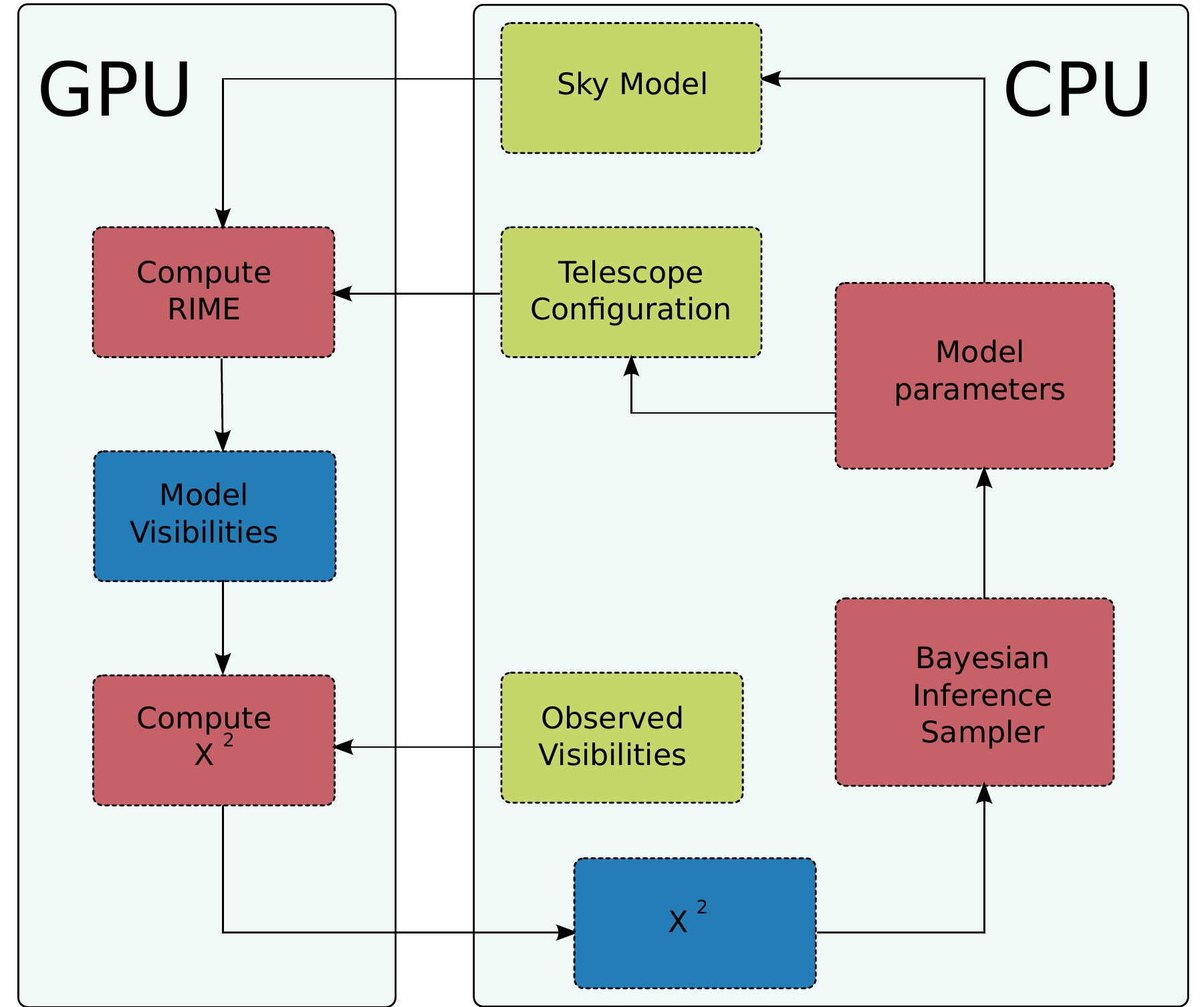}
\end{center}
\caption{BIRO algorithm flow. Red boxes indicate computation,
green boxes input and blue boxes output.
Given a parameterised sky model and a telescope configuration
defined in terms of direction-dependent and -independent effects,
the RIME is computed on the GPU to produce model visibilities.
These are used, in combination with observed visibilities
to produce a single, floating point, $\chi^2$ value. The $\chi^2$ is
transferred off the GPU onto the CPU and used by BIRO to estimate
new improved parameters for the sources in the sky model.
These parameters are uploaded to the GPU and the process is repeated
until the Bayesian inference stopping criterion is reached,
e.g. the goodness of fit indicated by the $\chi^2$ is sufficient.}
\label{fig:biro_workflow}
\end{figure}

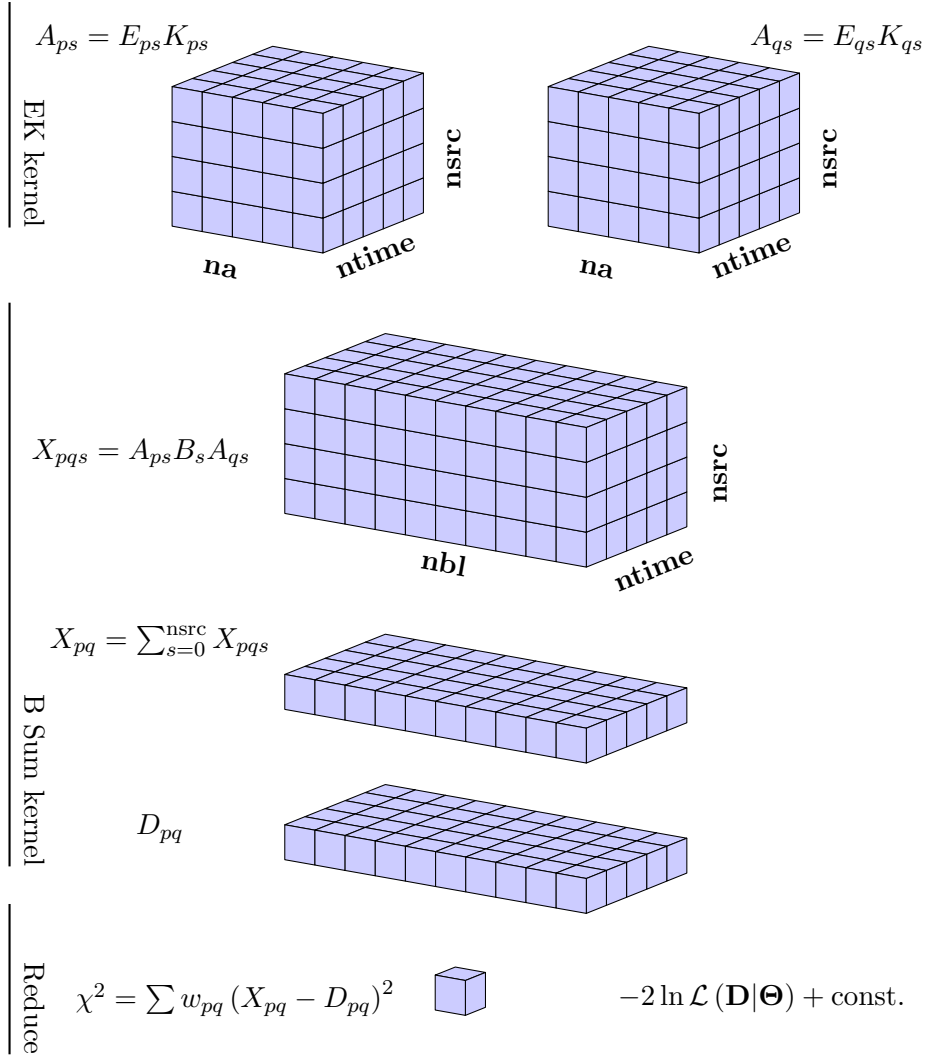
\begin{figure}
\begin{center}
% Sketch output, version 0.3 (build 7d, Wed May 2 06:36:52 2012)
% Output language: PGF/TikZ,LaTeX
\begin{tikzpicture}[line join=round]
\filldraw[thin,fill=blue!20!white](-2.004,-7.411)--(-2.272,-7.518)--(-1.871,-7.589)--(-1.604,-7.482)--cycle;
\filldraw[thin,fill=blue!20!white](-2.004,-3.411)--(-2.272,-3.518)--(-1.871,-3.589)--(-1.604,-3.482)--cycle;
\filldraw[thin,fill=blue!20!white](-2.004,-9.411)--(-2.272,-9.518)--(-1.871,-9.589)--(-1.604,-9.482)--cycle;
\filldraw[thin,fill=blue!20!white](-1.604,-7.482)--(-1.871,-7.589)--(-1.47,-7.661)--(-1.203,-7.554)--cycle;
\filldraw[thin,fill=blue!20!white](-1.604,-3.482)--(-1.871,-3.589)--(-1.47,-3.661)--(-1.203,-3.554)--cycle;
\filldraw[thin,fill=blue!20!white](-1.604,-9.482)--(-1.871,-9.589)--(-1.47,-9.661)--(-1.203,-9.554)--cycle;
\filldraw[thin,fill=blue!20!white](-2.272,-7.518)--(-2.539,-7.625)--(-2.138,-7.696)--(-1.871,-7.589)--cycle;
\filldraw[thin,fill=blue!20!white](-2.272,-9.518)--(-2.539,-9.625)--(-2.138,-9.696)--(-1.871,-9.589)--cycle;
\filldraw[thin,fill=blue!20!white](-2.272,-3.518)--(-2.539,-3.625)--(-2.138,-3.696)--(-1.871,-3.589)--cycle;
\filldraw[thin,fill=blue!20!white](-1.203,-7.554)--(-1.47,-7.661)--(-1.069,-7.732)--(-.802,-7.625)--cycle;
\filldraw[thin,fill=blue!20!white](-1.203,-3.554)--(-1.47,-3.661)--(-1.069,-3.732)--(-.802,-3.625)--cycle;
\filldraw[thin,fill=blue!20!white](-1.203,-9.554)--(-1.47,-9.661)--(-1.069,-9.732)--(-.802,-9.625)--cycle;
\filldraw[thin,fill=blue!20!white](-1.871,-9.589)--(-2.138,-9.696)--(-1.737,-9.768)--(-1.47,-9.661)--cycle;
\filldraw[thin,fill=blue!20!white](-1.871,-7.589)--(-2.138,-7.696)--(-1.737,-7.768)--(-1.47,-7.661)--cycle;
\filldraw[thin,fill=blue!20!white](-1.871,-3.589)--(-2.138,-3.696)--(-1.737,-3.768)--(-1.47,-3.661)--cycle;
\filldraw[thin,fill=blue!20!white](1.498,.411)--(1.231,.304)--(1.631,.232)--(1.899,.339)--cycle;
\filldraw[thin,fill=blue!20!white](-3.502,.411)--(-3.769,.304)--(-3.369,.232)--(-3.101,.339)--cycle;
\filldraw[thin,fill=blue!20!white](-2.539,-9.625)--(-2.806,-9.732)--(-2.405,-9.804)--(-2.138,-9.696)--cycle;
\filldraw[thin,fill=blue!20!white](-.802,-7.625)--(-1.069,-7.732)--(-.668,-7.804)--(-.401,-7.696)--cycle;
\filldraw[thin,fill=blue!20!white](-2.539,-7.625)--(-2.806,-7.732)--(-2.405,-7.804)--(-2.138,-7.696)--cycle;
\filldraw[thin,fill=blue!20!white](-.802,-3.625)--(-1.069,-3.732)--(-.668,-3.804)--(-.401,-3.696)--cycle;
\filldraw[thin,fill=blue!20!white](-2.539,-3.625)--(-2.806,-3.732)--(-2.405,-3.804)--(-2.138,-3.696)--cycle;
\filldraw[thin,fill=blue!20!white](-.802,-9.625)--(-1.069,-9.732)--(-.668,-9.804)--(-.401,-9.696)--cycle;
\filldraw[thin,fill=blue!20!white](1.899,.339)--(1.631,.232)--(2.032,.161)--(2.3,.268)--cycle;
\filldraw[thin,fill=blue!20!white](-3.101,.339)--(-3.369,.232)--(-2.968,.161)--(-2.7,.268)--cycle;
\filldraw[thin,fill=blue!20!white](-1.47,-7.661)--(-1.737,-7.768)--(-1.336,-7.839)--(-1.069,-7.732)--cycle;
\filldraw[thin,fill=blue!20!white](-1.47,-3.661)--(-1.737,-3.768)--(-1.336,-3.839)--(-1.069,-3.732)--cycle;
\filldraw[thin,fill=blue!20!white](-1.47,-9.661)--(-1.737,-9.768)--(-1.336,-9.839)--(-1.069,-9.732)--cycle;
\filldraw[thin,fill=blue!20!white](-.401,-9.696)--(-.668,-9.804)--(-.267,-9.875)--(0,-9.768)--cycle;
\filldraw[thin,fill=blue!20!white](-2.138,-9.696)--(-2.405,-9.804)--(-2.004,-9.875)--(-1.737,-9.768)--cycle;
\filldraw[thin,fill=blue!20!white](-.401,-7.696)--(-.668,-7.804)--(-.267,-7.875)--(0,-7.768)--cycle;
\filldraw[thin,fill=blue!20!white](-2.138,-7.696)--(-2.405,-7.804)--(-2.004,-7.875)--(-1.737,-7.768)--cycle;
\filldraw[thin,fill=blue!20!white](-.401,-3.696)--(-.668,-3.804)--(-.267,-3.875)--(0,-3.768)--cycle;
\filldraw[thin,fill=blue!20!white](-2.138,-3.696)--(-2.405,-3.804)--(-2.004,-3.875)--(-1.737,-3.768)--cycle;
\filldraw[thin,fill=blue!20!white](1.231,.304)--(.963,.196)--(1.364,.125)--(1.631,.232)--cycle;
\filldraw[thin,fill=blue!20!white](-3.769,.304)--(-4.037,.196)--(-3.636,.125)--(-3.369,.232)--cycle;
\filldraw[thin,fill=blue!20!white](2.3,.268)--(2.032,.161)--(2.433,.089)--(2.7,.196)--cycle;
\filldraw[thin,fill=blue!20!white](-2.806,-3.732)--(-3.074,-3.839)--(-2.673,-3.911)--(-2.405,-3.804)--cycle;
\filldraw[thin,fill=blue!20!white](-1.069,-3.732)--(-1.336,-3.839)--(-.935,-3.911)--(-.668,-3.804)--cycle;
\filldraw[thin,fill=blue!20!white](-2.806,-7.732)--(-3.074,-7.839)--(-2.673,-7.911)--(-2.405,-7.804)--cycle;
\filldraw[thin,fill=blue!20!white](-1.069,-7.732)--(-1.336,-7.839)--(-.935,-7.911)--(-.668,-7.804)--cycle;
\filldraw[thin,fill=blue!20!white](-2.806,-9.732)--(-3.074,-9.839)--(-2.673,-9.911)--(-2.405,-9.804)--cycle;
\filldraw[thin,fill=blue!20!white](-1.069,-9.732)--(-1.336,-9.839)--(-.935,-9.911)--(-.668,-9.804)--cycle;
\filldraw[thin,fill=blue!20!white](-2.7,.268)--(-2.968,.161)--(-2.567,.089)--(-2.3,.196)--cycle;
\filldraw[thin,fill=blue!20!white](-3.369,.232)--(-3.636,.125)--(-3.235,.054)--(-2.968,.161)--cycle;
\filldraw[thin,fill=blue!20!white](1.631,.232)--(1.364,.125)--(1.765,.054)--(2.032,.161)--cycle;
\filldraw[thin,fill=blue!20!white](0,-3.768)--(-.267,-3.875)--(.134,-3.946)--(.401,-3.839)--cycle;
\filldraw[thin,fill=blue!20!white](0,-7.768)--(-.267,-7.875)--(.134,-7.946)--(.401,-7.839)--cycle;
\filldraw[thin,fill=blue!20!white](0,-9.768)--(-.267,-9.875)--(.134,-9.946)--(.401,-9.839)--cycle;
\filldraw[thin,fill=blue!20!white](-1.737,-7.768)--(-2.004,-7.875)--(-1.604,-7.946)--(-1.336,-7.839)--cycle;
\filldraw[thin,fill=blue!20!white](-1.737,-3.768)--(-2.004,-3.875)--(-1.604,-3.946)--(-1.336,-3.839)--cycle;
\filldraw[thin,fill=blue!20!white](-1.737,-9.768)--(-2.004,-9.875)--(-1.604,-9.946)--(-1.336,-9.839)--cycle;
\filldraw[thin,fill=blue!20!white](-.668,-9.804)--(-.935,-9.911)--(-.535,-9.982)--(-.267,-9.875)--cycle;
\filldraw[thin,fill=blue!20!white](-2.405,-9.804)--(-2.673,-9.911)--(-2.272,-9.982)--(-2.004,-9.875)--cycle;
\filldraw[thin,fill=blue!20!white](-.668,-7.804)--(-.935,-7.911)--(-.535,-7.982)--(-.267,-7.875)--cycle;
\filldraw[thin,fill=blue!20!white](-2.405,-7.804)--(-2.673,-7.911)--(-2.272,-7.982)--(-2.004,-7.875)--cycle;
\filldraw[thin,fill=blue!20!white](-.668,-3.804)--(-.935,-3.911)--(-.535,-3.982)--(-.267,-3.875)--cycle;
\filldraw[thin,fill=blue!20!white](-2.405,-3.804)--(-2.673,-3.911)--(-2.272,-3.982)--(-2.004,-3.875)--cycle;
\filldraw[thin,fill=blue!20!white](-3.341,-5.804)--(-2.94,-5.875)--(-2.94,-5.411)--(-3.341,-5.339)--cycle;
\filldraw[thin,fill=blue!20!white](2.7,.196)--(2.433,.089)--(2.834,.018)--(3.101,.125)--cycle;
\filldraw[thin,fill=blue!20!white](.963,.196)--(.696,.089)--(1.097,.018)--(1.364,.125)--cycle;
\filldraw[thin,fill=blue!20!white](3.502,-1.804)--(3.502,-1.339)--(3.235,-1.446)--(3.235,-1.911)--cycle;
\filldraw[thin,fill=blue!20!white](-2.3,.196)--(-2.567,.089)--(-2.166,.018)--(-1.899,.125)--cycle;
\filldraw[thin,fill=blue!20!white](-4.037,.196)--(-4.304,.089)--(-3.903,.018)--(-3.636,.125)--cycle;
\filldraw[thin,fill=blue!20!white](-1.498,-1.804)--(-1.498,-1.339)--(-1.765,-1.446)--(-1.765,-1.911)--cycle;
\filldraw[thin,fill=blue!20!white](.401,-9.839)--(.134,-9.946)--(.535,-10.018)--(.802,-9.911)--cycle;
\filldraw[thin,fill=blue!20!white](-1.336,-9.839)--(-1.604,-9.946)--(-1.203,-10.018)--(-.935,-9.911)--cycle;
\filldraw[thin,fill=blue!20!white](-3.074,-9.839)--(-3.341,-9.946)--(-2.94,-10.018)--(-2.673,-9.911)--cycle;
\filldraw[thin,fill=blue!20!white](.401,-7.839)--(.134,-7.946)--(.535,-8.018)--(.802,-7.911)--cycle;
\filldraw[thin,fill=blue!20!white](-1.336,-7.839)--(-1.604,-7.946)--(-1.203,-8.018)--(-.935,-7.911)--cycle;
\filldraw[thin,fill=blue!20!white](-3.074,-7.839)--(-3.341,-7.946)--(-2.94,-8.018)--(-2.673,-7.911)--cycle;
\filldraw[thin,fill=blue!20!white](.401,-3.839)--(.134,-3.946)--(.535,-4.018)--(.802,-3.911)--cycle;
\filldraw[thin,fill=blue!20!white](-1.336,-3.839)--(-1.604,-3.946)--(-1.203,-4.018)--(-.935,-3.911)--cycle;
\filldraw[thin,fill=blue!20!white](-3.074,-3.839)--(-3.341,-3.946)--(-2.94,-4.018)--(-2.673,-3.911)--cycle;
\filldraw[thin,fill=blue!20!white](-3.341,-5.339)--(-2.94,-5.411)--(-2.94,-4.946)--(-3.341,-4.875)--cycle;
\filldraw[thin,fill=blue!20!white](2.032,.161)--(1.765,.054)--(2.166,-.018)--(2.433,.089)--cycle;
\filldraw[thin,fill=blue!20!white](3.502,-1.339)--(3.502,-.875)--(3.235,-.982)--(3.235,-1.446)--cycle;
\filldraw[thin,fill=blue!20!white](-2.968,.161)--(-3.235,.054)--(-2.834,-.018)--(-2.567,.089)--cycle;
\filldraw[thin,fill=blue!20!white](-1.498,-1.339)--(-1.498,-.875)--(-1.765,-.982)--(-1.765,-1.446)--cycle;
\filldraw[thin,fill=blue!20!white](-1.074,-11.839)--(-1.341,-11.946)--(-.94,-12.018)--(-.673,-11.911)--cycle;
\filldraw[thin,fill=blue!20!white](-.267,-9.875)--(-.535,-9.982)--(-.134,-10.054)--(.134,-9.946)--cycle;
\filldraw[thin,fill=blue!20!white](-.267,-7.875)--(-.535,-7.982)--(-.134,-8.054)--(.134,-7.946)--cycle;
\filldraw[thin,fill=blue!20!white](-.267,-3.875)--(-.535,-3.982)--(-.134,-4.054)--(.134,-3.946)--cycle;
\filldraw[thin,fill=blue!20!white](-2.94,-5.875)--(-2.539,-5.946)--(-2.539,-5.482)--(-2.94,-5.411)--cycle;
\filldraw[thin,fill=blue!20!white](-3.341,-4.875)--(-2.94,-4.946)--(-2.94,-4.482)--(-3.341,-4.411)--cycle;
\filldraw[thin,fill=blue!20!white](3.101,.125)--(2.834,.018)--(3.235,-.054)--(3.502,.054)--cycle;
\filldraw[thin,fill=blue!20!white](1.364,.125)--(1.097,.018)--(1.498,-.054)--(1.765,.054)--cycle;
\filldraw[thin,fill=blue!20!white](3.502,-.875)--(3.502,-.411)--(3.235,-.518)--(3.235,-.982)--cycle;
\filldraw[thin,fill=blue!20!white](-1.899,.125)--(-2.166,.018)--(-1.765,-.054)--(-1.498,.054)--cycle;
\filldraw[thin,fill=blue!20!white](-3.636,.125)--(-3.903,.018)--(-3.502,-.054)--(-3.235,.054)--cycle;
\filldraw[thin,fill=blue!20!white](-1.498,-.875)--(-1.498,-.411)--(-1.765,-.518)--(-1.765,-.982)--cycle;
\filldraw[thin,fill=blue!20!white](-.673,-12.375)--(-.673,-11.911)--(-.94,-12.018)--(-.94,-12.482)--cycle;
\filldraw[thin,fill=blue!20!white](-2.004,-7.875)--(-2.272,-7.982)--(-1.871,-8.054)--(-1.604,-7.946)--cycle;
\filldraw[thin,fill=blue!20!white](-2.004,-3.875)--(-2.272,-3.982)--(-1.871,-4.054)--(-1.604,-3.946)--cycle;
\filldraw[thin,fill=blue!20!white](-2.004,-9.875)--(-2.272,-9.982)--(-1.871,-10.054)--(-1.604,-9.946)--cycle;
\filldraw[thin,fill=blue!20!white](-1.341,-12.411)--(-.94,-12.482)--(-.94,-12.018)--(-1.341,-11.946)--cycle;
\filldraw[thin,fill=blue!20!white](.802,-9.911)--(.535,-10.018)--(.935,-10.089)--(1.203,-9.982)--cycle;
\filldraw[thin,fill=blue!20!white](-.935,-9.911)--(-1.203,-10.018)--(-.802,-10.089)--(-.535,-9.982)--cycle;
\filldraw[thin,fill=blue!20!white](-2.673,-9.911)--(-2.94,-10.018)--(-2.539,-10.089)--(-2.272,-9.982)--cycle;
\filldraw[thin,fill=blue!20!white](-3.341,-10.411)--(-2.94,-10.482)--(-2.94,-10.018)--(-3.341,-9.946)--cycle;
\filldraw[thin,fill=blue!20!white](.802,-7.911)--(.535,-8.018)--(.935,-8.089)--(1.203,-7.982)--cycle;
\filldraw[thin,fill=blue!20!white](-.935,-7.911)--(-1.203,-8.018)--(-.802,-8.089)--(-.535,-7.982)--cycle;
\filldraw[thin,fill=blue!20!white](-2.673,-7.911)--(-2.94,-8.018)--(-2.539,-8.089)--(-2.272,-7.982)--cycle;
\filldraw[thin,fill=blue!20!white](-3.341,-8.411)--(-2.94,-8.482)--(-2.94,-8.018)--(-3.341,-7.946)--cycle;
\filldraw[thin,fill=blue!20!white](.802,-3.911)--(.535,-4.018)--(.935,-4.089)--(1.203,-3.982)--cycle;
\filldraw[thin,fill=blue!20!white](-.935,-3.911)--(-1.203,-4.018)--(-.802,-4.089)--(-.535,-3.982)--cycle;
\filldraw[thin,fill=blue!20!white](-2.673,-3.911)--(-2.94,-4.018)--(-2.539,-4.089)--(-2.272,-3.982)--cycle;
\filldraw[thin,fill=blue!20!white](-2.94,-5.411)--(-2.539,-5.482)--(-2.539,-5.018)--(-2.94,-4.946)--cycle;
\filldraw[thin,fill=blue!20!white](-3.341,-4.411)--(-2.94,-4.482)--(-2.94,-4.018)--(-3.341,-3.946)--cycle;
\filldraw[thin,fill=blue!20!white](2.433,.089)--(2.166,-.018)--(2.567,-.089)--(2.834,.018)--cycle;
\filldraw[thin,fill=blue!20!white](.696,.089)--(.429,-.018)--(.83,-.089)--(1.097,.018)--cycle;
\filldraw[thin,fill=blue!20!white](3.235,-1.911)--(3.235,-1.446)--(2.968,-1.554)--(2.968,-2.018)--cycle;
\filldraw[thin,fill=blue!20!white](3.502,-.411)--(3.502,.054)--(3.235,-.054)--(3.235,-.518)--cycle;
\filldraw[thin,fill=blue!20!white](-2.567,.089)--(-2.834,-.018)--(-2.433,-.089)--(-2.166,.018)--cycle;
\filldraw[thin,fill=blue!20!white](-4.304,.089)--(-4.571,-.018)--(-4.17,-.089)--(-3.903,.018)--cycle;
\filldraw[thin,fill=blue!20!white](-1.765,-1.911)--(-1.765,-1.446)--(-2.032,-1.554)--(-2.032,-2.018)--cycle;
\filldraw[thin,fill=blue!20!white](-1.498,-.411)--(-1.498,.054)--(-1.765,-.054)--(-1.765,-.518)--cycle;
\filldraw[thin,fill=blue!20!white](-2.94,-4.946)--(-2.539,-5.018)--(-2.539,-4.554)--(-2.94,-4.482)--cycle;
\filldraw[thin,fill=blue!20!white](.134,-9.946)--(-.134,-10.054)--(.267,-10.125)--(.535,-10.018)--cycle;
\filldraw[thin,fill=blue!20!white](-1.604,-9.946)--(-1.871,-10.054)--(-1.47,-10.125)--(-1.203,-10.018)--cycle;
\filldraw[thin,fill=blue!20!white](.134,-7.946)--(-.134,-8.054)--(.267,-8.125)--(.535,-8.018)--cycle;
\filldraw[thin,fill=blue!20!white](-1.604,-7.946)--(-1.871,-8.054)--(-1.47,-8.125)--(-1.203,-8.018)--cycle;
\filldraw[thin,fill=blue!20!white](.134,-3.946)--(-.134,-4.054)--(.267,-4.125)--(.535,-4.018)--cycle;
\filldraw[thin,fill=blue!20!white](-1.604,-3.946)--(-1.871,-4.054)--(-1.47,-4.125)--(-1.203,-4.018)--cycle;
\filldraw[thin,fill=blue!20!white](-2.539,-5.946)--(-2.138,-6.018)--(-2.138,-5.554)--(-2.539,-5.482)--cycle;
\filldraw[thin,fill=blue!20!white](1.765,.054)--(1.498,-.054)--(1.899,-.125)--(2.166,-.018)--cycle;
\filldraw[thin,fill=blue!20!white](3.235,-1.446)--(3.235,-.982)--(2.968,-1.089)--(2.968,-1.554)--cycle;
\filldraw[thin,fill=blue!20!white](-3.235,.054)--(-3.502,-.054)--(-3.101,-.125)--(-2.834,-.018)--cycle;
\filldraw[thin,fill=blue!20!white](-1.765,-1.446)--(-1.765,-.982)--(-2.032,-1.089)--(-2.032,-1.554)--cycle;
\filldraw[thin,fill=blue!20!white](-1.765,-.982)--(-1.765,-.518)--(-2.032,-.625)--(-2.032,-1.089)--cycle;
\filldraw[thin,fill=blue!20!white](-4.839,-1.982)--(-4.438,-2.054)--(-4.438,-1.589)--(-4.839,-1.518)--cycle;
\filldraw[thin,fill=blue!20!white](-3.903,.018)--(-4.17,-.089)--(-3.769,-.161)--(-3.502,-.054)--cycle;
\filldraw[thin,fill=blue!20!white](-2.166,.018)--(-2.433,-.089)--(-2.032,-.161)--(-1.765,-.054)--cycle;
\filldraw[thin,fill=blue!20!white](3.235,-.982)--(3.235,-.518)--(2.968,-.625)--(2.968,-1.089)--cycle;
\filldraw[thin,fill=blue!20!white](.161,-1.982)--(.562,-2.054)--(.562,-1.589)--(.161,-1.518)--cycle;
\filldraw[thin,fill=blue!20!white](1.097,.018)--(.83,-.089)--(1.231,-.161)--(1.498,-.054)--cycle;
\filldraw[thin,fill=blue!20!white](2.834,.018)--(2.567,-.089)--(2.968,-.161)--(3.235,-.054)--cycle;
\filldraw[thin,fill=blue!20!white](-2.94,-4.482)--(-2.539,-4.554)--(-2.539,-4.089)--(-2.94,-4.018)--cycle;
\filldraw[thin,fill=blue!20!white](2.004,-5.982)--(2.004,-5.518)--(1.737,-5.625)--(1.737,-6.089)--cycle;
\filldraw[thin,fill=blue!20!white](-.535,-3.982)--(-.802,-4.089)--(-.401,-4.161)--(-.134,-4.054)--cycle;
\filldraw[thin,fill=blue!20!white](1.203,-3.982)--(.935,-4.089)--(1.336,-4.161)--(1.604,-4.054)--cycle;
\filldraw[thin,fill=blue!20!white](-2.94,-8.482)--(-2.539,-8.554)--(-2.539,-8.089)--(-2.94,-8.018)--cycle;
\filldraw[thin,fill=blue!20!white](-.535,-7.982)--(-.802,-8.089)--(-.401,-8.161)--(-.134,-8.054)--cycle;
\filldraw[thin,fill=blue!20!white](1.203,-7.982)--(.935,-8.089)--(1.336,-8.161)--(1.604,-8.054)--cycle;
\filldraw[thin,fill=blue!20!white](-2.94,-10.482)--(-2.539,-10.554)--(-2.539,-10.089)--(-2.94,-10.018)--cycle;
\filldraw[thin,fill=blue!20!white](-.535,-9.982)--(-.802,-10.089)--(-.401,-10.161)--(-.134,-10.054)--cycle;
\filldraw[thin,fill=blue!20!white](1.203,-9.982)--(.935,-10.089)--(1.336,-10.161)--(1.604,-10.054)--cycle;
\filldraw[thin,fill=blue!20!white](-2.272,-7.982)--(-2.539,-8.089)--(-2.138,-8.161)--(-1.871,-8.054)--cycle;
\filldraw[thin,fill=blue!20!white](-2.272,-3.982)--(-2.539,-4.089)--(-2.138,-4.161)--(-1.871,-4.054)--cycle;
\filldraw[thin,fill=blue!20!white](-2.539,-5.482)--(-2.138,-5.554)--(-2.138,-5.089)--(-2.539,-5.018)--cycle;
\filldraw[thin,fill=blue!20!white](-2.272,-9.982)--(-2.539,-10.089)--(-2.138,-10.161)--(-1.871,-10.054)--cycle;
\filldraw[thin,fill=blue!20!white](.535,-10.018)--(.267,-10.125)--(.668,-10.196)--(.935,-10.089)--cycle;
\filldraw[thin,fill=blue!20!white](-1.203,-10.018)--(-1.47,-10.125)--(-1.069,-10.196)--(-.802,-10.089)--cycle;
\filldraw[thin,fill=blue!20!white](.535,-8.018)--(.267,-8.125)--(.668,-8.196)--(.935,-8.089)--cycle;
\filldraw[thin,fill=blue!20!white](-1.203,-8.018)--(-1.47,-8.125)--(-1.069,-8.196)--(-.802,-8.089)--cycle;
\filldraw[thin,fill=blue!20!white](.535,-4.018)--(.267,-4.125)--(.668,-4.196)--(.935,-4.089)--cycle;
\filldraw[thin,fill=blue!20!white](-1.203,-4.018)--(-1.47,-4.125)--(-1.069,-4.196)--(-.802,-4.089)--cycle;
\filldraw[thin,fill=blue!20!white](-2.138,-6.018)--(-1.737,-6.089)--(-1.737,-5.625)--(-2.138,-5.554)--cycle;
\filldraw[thin,fill=blue!20!white](2.004,-5.518)--(2.004,-5.054)--(1.737,-5.161)--(1.737,-5.625)--cycle;
\filldraw[thin,fill=blue!20!white](-2.539,-5.018)--(-2.138,-5.089)--(-2.138,-4.625)--(-2.539,-4.554)--cycle;
\filldraw[thin,fill=blue!20!white](2.166,-.018)--(1.899,-.125)--(2.3,-.196)--(2.567,-.089)--cycle;
\filldraw[thin,fill=blue!20!white](.429,-.018)--(.161,-.125)--(.562,-.196)--(.83,-.089)--cycle;
\filldraw[thin,fill=blue!20!white](2.968,-2.018)--(2.968,-1.554)--(2.7,-1.661)--(2.7,-2.125)--cycle;
\filldraw[thin,fill=blue!20!white](.161,-1.518)--(.562,-1.589)--(.562,-1.125)--(.161,-1.054)--cycle;
\filldraw[thin,fill=blue!20!white](3.235,-.518)--(3.235,-.054)--(2.968,-.161)--(2.968,-.625)--cycle;
\filldraw[thin,fill=blue!20!white](-2.834,-.018)--(-3.101,-.125)--(-2.7,-.196)--(-2.433,-.089)--cycle;
\filldraw[thin,fill=blue!20!white](-4.571,-.018)--(-4.839,-.125)--(-4.438,-.196)--(-4.17,-.089)--cycle;
\filldraw[thin,fill=blue!20!white](-2.032,-2.018)--(-2.032,-1.554)--(-2.3,-1.661)--(-2.3,-2.125)--cycle;
\filldraw[thin,fill=blue!20!white](-4.839,-1.518)--(-4.438,-1.589)--(-4.438,-1.125)--(-4.839,-1.054)--cycle;
\filldraw[thin,fill=blue!20!white](-1.765,-.518)--(-1.765,-.054)--(-2.032,-.161)--(-2.032,-.625)--cycle;
\filldraw[thin,fill=blue!20!white](-4.438,-2.054)--(-4.037,-2.125)--(-4.037,-1.661)--(-4.438,-1.589)--cycle;
\filldraw[thin,fill=blue!20!white](.161,-1.054)--(.562,-1.125)--(.562,-.661)--(.161,-.589)--cycle;
\filldraw[thin,fill=blue!20!white](.562,-2.054)--(.963,-2.125)--(.963,-1.661)--(.562,-1.589)--cycle;
\filldraw[thin,fill=blue!20!white](2.004,-5.054)--(2.004,-4.589)--(1.737,-4.696)--(1.737,-5.161)--cycle;
\filldraw[thin,fill=blue!20!white](-4.839,-1.054)--(-4.438,-1.125)--(-4.438,-.661)--(-4.839,-.589)--cycle;
\filldraw[thin,fill=blue!20!white](-2.032,-1.554)--(-2.032,-1.089)--(-2.3,-1.196)--(-2.3,-1.661)--cycle;
\filldraw[thin,fill=blue!20!white](-3.502,-.054)--(-3.769,-.161)--(-3.369,-.232)--(-3.101,-.125)--cycle;
\filldraw[thin,fill=blue!20!white](2.968,-1.554)--(2.968,-1.089)--(2.7,-1.196)--(2.7,-1.661)--cycle;
\filldraw[thin,fill=blue!20!white](1.498,-.054)--(1.231,-.161)--(1.631,-.232)--(1.899,-.125)--cycle;
\filldraw[thin,fill=blue!20!white](-2.539,-4.554)--(-2.138,-4.625)--(-2.138,-4.161)--(-2.539,-4.089)--cycle;
\filldraw[thin,fill=blue!20!white](-2.138,-5.554)--(-1.737,-5.625)--(-1.737,-5.161)--(-2.138,-5.089)--cycle;
\filldraw[thin,fill=blue!20!white](-1.871,-4.054)--(-2.138,-4.161)--(-1.737,-4.232)--(-1.47,-4.125)--cycle;
\filldraw[thin,fill=blue!20!white](-.134,-4.054)--(-.401,-4.161)--(0,-4.232)--(.267,-4.125)--cycle;
\filldraw[thin,fill=blue!20!white](1.604,-4.054)--(1.336,-4.161)--(1.737,-4.232)--(2.004,-4.125)--cycle;
\filldraw[thin,fill=blue!20!white](-2.539,-8.554)--(-2.138,-8.625)--(-2.138,-8.161)--(-2.539,-8.089)--cycle;
\filldraw[thin,fill=blue!20!white](-1.871,-8.054)--(-2.138,-8.161)--(-1.737,-8.232)--(-1.47,-8.125)--cycle;
\filldraw[thin,fill=blue!20!white](-.134,-8.054)--(-.401,-8.161)--(0,-8.232)--(.267,-8.125)--cycle;
\filldraw[thin,fill=blue!20!white](1.604,-8.054)--(1.336,-8.161)--(1.737,-8.232)--(2.004,-8.125)--cycle;
\filldraw[thin,fill=blue!20!white](-2.539,-10.554)--(-2.138,-10.625)--(-2.138,-10.161)--(-2.539,-10.089)--cycle;
\filldraw[thin,fill=blue!20!white](-1.871,-10.054)--(-2.138,-10.161)--(-1.737,-10.232)--(-1.47,-10.125)--cycle;
\filldraw[thin,fill=blue!20!white](-.134,-10.054)--(-.401,-10.161)--(0,-10.232)--(.267,-10.125)--cycle;
\filldraw[thin,fill=blue!20!white](1.604,-10.054)--(1.336,-10.161)--(1.737,-10.232)--(2.004,-10.125)--cycle;
\filldraw[thin,fill=blue!20!white](.935,-10.089)--(.668,-10.196)--(1.069,-10.268)--(1.336,-10.161)--cycle;
\filldraw[thin,fill=blue!20!white](-.802,-10.089)--(-1.069,-10.196)--(-.668,-10.268)--(-.401,-10.161)--cycle;
\filldraw[thin,fill=blue!20!white](2.004,-10.589)--(2.004,-10.125)--(1.737,-10.232)--(1.737,-10.696)--cycle;
\filldraw[thin,fill=blue!20!white](.935,-8.089)--(.668,-8.196)--(1.069,-8.268)--(1.336,-8.161)--cycle;
\filldraw[thin,fill=blue!20!white](-.802,-8.089)--(-1.069,-8.196)--(-.668,-8.268)--(-.401,-8.161)--cycle;
\filldraw[thin,fill=blue!20!white](2.004,-8.589)--(2.004,-8.125)--(1.737,-8.232)--(1.737,-8.696)--cycle;
\filldraw[thin,fill=blue!20!white](.935,-4.089)--(.668,-4.196)--(1.069,-4.268)--(1.336,-4.161)--cycle;
\filldraw[thin,fill=blue!20!white](-.802,-4.089)--(-1.069,-4.196)--(-.668,-4.268)--(-.401,-4.161)--cycle;
\filldraw[thin,fill=blue!20!white](1.737,-6.089)--(1.737,-5.625)--(1.47,-5.732)--(1.47,-6.196)--cycle;
\filldraw[thin,fill=blue!20!white](-1.737,-6.089)--(-1.336,-6.161)--(-1.336,-5.696)--(-1.737,-5.625)--cycle;
\filldraw[thin,fill=blue!20!white](-2.138,-5.089)--(-1.737,-5.161)--(-1.737,-4.696)--(-2.138,-4.625)--cycle;
\filldraw[thin,fill=blue!20!white](2.004,-4.589)--(2.004,-4.125)--(1.737,-4.232)--(1.737,-4.696)--cycle;
\filldraw[thin,fill=blue!20!white](2.567,-.089)--(2.3,-.196)--(2.7,-.268)--(2.968,-.161)--cycle;
\filldraw[thin,fill=blue!20!white](.83,-.089)--(.562,-.196)--(.963,-.268)--(1.231,-.161)--cycle;
\filldraw[thin,fill=blue!20!white](.562,-1.589)--(.963,-1.661)--(.963,-1.196)--(.562,-1.125)--cycle;
\filldraw[thin,fill=blue!20!white](2.968,-1.089)--(2.968,-.625)--(2.7,-.732)--(2.7,-1.196)--cycle;
\filldraw[thin,fill=blue!20!white](.161,-.589)--(.562,-.661)--(.562,-.196)--(.161,-.125)--cycle;
\filldraw[thin,fill=blue!20!white](-2.433,-.089)--(-2.7,-.196)--(-2.3,-.268)--(-2.032,-.161)--cycle;
\filldraw[thin,fill=blue!20!white](-4.17,-.089)--(-4.438,-.196)--(-4.037,-.268)--(-3.769,-.161)--cycle;
\filldraw[thin,fill=blue!20!white](-4.438,-1.589)--(-4.037,-1.661)--(-4.037,-1.196)--(-4.438,-1.125)--cycle;
\filldraw[thin,fill=blue!20!white](-2.032,-1.089)--(-2.032,-.625)--(-2.3,-.732)--(-2.3,-1.196)--cycle;
\filldraw[thin,fill=blue!20!white](-4.839,-.589)--(-4.438,-.661)--(-4.438,-.196)--(-4.839,-.125)--cycle;
\filldraw[thin,fill=blue!20!white](-2.3,-2.125)--(-2.3,-1.661)--(-2.567,-1.768)--(-2.567,-2.232)--cycle;
\filldraw[thin,fill=blue!20!white](.562,-1.125)--(.963,-1.196)--(.963,-.732)--(.562,-.661)--cycle;
\filldraw[thin,fill=blue!20!white](2.7,-2.125)--(2.7,-1.661)--(2.433,-1.768)--(2.433,-2.232)--cycle;
\filldraw[thin,fill=blue!20!white](-4.438,-1.125)--(-4.037,-1.196)--(-4.037,-.732)--(-4.438,-.661)--cycle;
\filldraw[thin,fill=blue!20!white](-1.47,-10.125)--(-1.737,-10.232)--(-1.336,-10.304)--(-1.069,-10.196)--cycle;
\filldraw[thin,fill=blue!20!white](-2.138,-10.625)--(-1.737,-10.696)--(-1.737,-10.232)--(-2.138,-10.161)--cycle;
\filldraw[thin,fill=blue!20!white](.267,-8.125)--(0,-8.232)--(.401,-8.304)--(.668,-8.196)--cycle;
\filldraw[thin,fill=blue!20!white](-1.47,-8.125)--(-1.737,-8.232)--(-1.336,-8.304)--(-1.069,-8.196)--cycle;
\filldraw[thin,fill=blue!20!white](-2.138,-8.625)--(-1.737,-8.696)--(-1.737,-8.232)--(-2.138,-8.161)--cycle;
\filldraw[thin,fill=blue!20!white](.267,-4.125)--(0,-4.232)--(.401,-4.304)--(.668,-4.196)--cycle;
\filldraw[thin,fill=blue!20!white](-1.47,-4.125)--(-1.737,-4.232)--(-1.336,-4.304)--(-1.069,-4.196)--cycle;
\filldraw[thin,fill=blue!20!white](1.737,-5.625)--(1.737,-5.161)--(1.47,-5.268)--(1.47,-5.732)--cycle;
\filldraw[thin,fill=blue!20!white](-1.737,-5.625)--(-1.336,-5.696)--(-1.336,-5.232)--(-1.737,-5.161)--cycle;
\filldraw[thin,fill=blue!20!white](-2.138,-4.625)--(-1.737,-4.696)--(-1.737,-4.232)--(-2.138,-4.161)--cycle;
\filldraw[thin,fill=blue!20!white](1.899,-.125)--(1.631,-.232)--(2.032,-.304)--(2.3,-.196)--cycle;
\filldraw[thin,fill=blue!20!white](.963,-2.125)--(1.364,-2.196)--(1.364,-1.732)--(.963,-1.661)--cycle;
\filldraw[thin,fill=blue!20!white](2.968,-.625)--(2.968,-.161)--(2.7,-.268)--(2.7,-.732)--cycle;
\filldraw[thin,fill=blue!20!white](-3.101,-.125)--(-3.369,-.232)--(-2.968,-.304)--(-2.7,-.196)--cycle;
\filldraw[thin,fill=blue!20!white](-4.037,-2.125)--(-3.636,-2.196)--(-3.636,-1.732)--(-4.037,-1.661)--cycle;
\filldraw[thin,fill=blue!20!white](-2.032,-.625)--(-2.032,-.161)--(-2.3,-.268)--(-2.3,-.732)--cycle;
\filldraw[thin,fill=blue!20!white](.267,-10.125)--(0,-10.232)--(.401,-10.304)--(.668,-10.196)--cycle;
\filldraw[thin,fill=blue!20!white](.562,-.661)--(.963,-.732)--(.963,-.268)--(.562,-.196)--cycle;
\filldraw[thin,fill=blue!20!white](-1.737,-5.161)--(-1.336,-5.232)--(-1.336,-4.768)--(-1.737,-4.696)--cycle;
\filldraw[thin,fill=blue!20!white](1.737,-5.161)--(1.737,-4.696)--(1.47,-4.804)--(1.47,-5.268)--cycle;
\filldraw[thin,fill=blue!20!white](-1.336,-6.161)--(-.935,-6.232)--(-.935,-5.768)--(-1.336,-5.696)--cycle;
\filldraw[thin,fill=blue!20!white](-4.438,-.661)--(-4.037,-.732)--(-4.037,-.268)--(-4.438,-.196)--cycle;
\filldraw[thin,fill=blue!20!white](-4.037,-1.661)--(-3.636,-1.732)--(-3.636,-1.268)--(-4.037,-1.196)--cycle;
\filldraw[thin,fill=blue!20!white](-2.3,-1.661)--(-2.3,-1.196)--(-2.567,-1.304)--(-2.567,-1.768)--cycle;
\filldraw[thin,fill=blue!20!white](-3.769,-.161)--(-4.037,-.268)--(-3.636,-.339)--(-3.369,-.232)--cycle;
\filldraw[thin,fill=blue!20!white](.963,-1.661)--(1.364,-1.732)--(1.364,-1.268)--(.963,-1.196)--cycle;
\filldraw[thin,fill=blue!20!white](2.7,-1.661)--(2.7,-1.196)--(2.433,-1.304)--(2.433,-1.768)--cycle;
\filldraw[thin,fill=blue!20!white](1.231,-.161)--(.963,-.268)--(1.364,-.339)--(1.631,-.232)--cycle;
\filldraw[thin,fill=blue!20!white](-.401,-4.161)--(-.668,-4.268)--(-.267,-4.339)--(0,-4.232)--cycle;
\filldraw[thin,fill=blue!20!white](1.336,-4.161)--(1.069,-4.268)--(1.47,-4.339)--(1.737,-4.232)--cycle;
\filldraw[thin,fill=blue!20!white](-.401,-8.161)--(-.668,-8.268)--(-.267,-8.339)--(0,-8.232)--cycle;
\filldraw[thin,fill=blue!20!white](1.336,-8.161)--(1.069,-8.268)--(1.47,-8.339)--(1.737,-8.232)--cycle;
\filldraw[thin,fill=blue!20!white](-.401,-10.161)--(-.668,-10.268)--(-.267,-10.339)--(0,-10.232)--cycle;
\filldraw[thin,fill=blue!20!white](1.336,-10.161)--(1.069,-10.268)--(1.47,-10.339)--(1.737,-10.232)--cycle;
\filldraw[thin,fill=blue!20!white](-1.069,-10.196)--(-1.336,-10.304)--(-.935,-10.375)--(-.668,-10.268)--cycle;
\filldraw[thin,fill=blue!20!white](1.737,-10.696)--(1.737,-10.232)--(1.47,-10.339)--(1.47,-10.804)--cycle;
\filldraw[thin,fill=blue!20!white](-1.737,-10.696)--(-1.336,-10.768)--(-1.336,-10.304)--(-1.737,-10.232)--cycle;
\filldraw[thin,fill=blue!20!white](.668,-8.196)--(.401,-8.304)--(.802,-8.375)--(1.069,-8.268)--cycle;
\filldraw[thin,fill=blue!20!white](-1.069,-8.196)--(-1.336,-8.304)--(-.935,-8.375)--(-.668,-8.268)--cycle;
\filldraw[thin,fill=blue!20!white](1.737,-8.696)--(1.737,-8.232)--(1.47,-8.339)--(1.47,-8.804)--cycle;
\filldraw[thin,fill=blue!20!white](-1.737,-8.696)--(-1.336,-8.768)--(-1.336,-8.304)--(-1.737,-8.232)--cycle;
\filldraw[thin,fill=blue!20!white](.668,-4.196)--(.401,-4.304)--(.802,-4.375)--(1.069,-4.268)--cycle;
\filldraw[thin,fill=blue!20!white](-1.069,-4.196)--(-1.336,-4.304)--(-.935,-4.375)--(-.668,-4.268)--cycle;
\filldraw[thin,fill=blue!20!white](1.47,-6.196)--(1.47,-5.732)--(1.203,-5.839)--(1.203,-6.304)--cycle;
\filldraw[thin,fill=blue!20!white](-1.336,-5.696)--(-.935,-5.768)--(-.935,-5.304)--(-1.336,-5.232)--cycle;
\filldraw[thin,fill=blue!20!white](1.737,-4.696)--(1.737,-4.232)--(1.47,-4.339)--(1.47,-4.804)--cycle;
\filldraw[thin,fill=blue!20!white](-1.737,-4.696)--(-1.336,-4.768)--(-1.336,-4.304)--(-1.737,-4.232)--cycle;
\filldraw[thin,fill=blue!20!white](2.3,-.196)--(2.032,-.304)--(2.433,-.375)--(2.7,-.268)--cycle;
\filldraw[thin,fill=blue!20!white](1.364,-2.196)--(1.765,-2.268)--(1.765,-1.804)--(1.364,-1.732)--cycle;
\filldraw[thin,fill=blue!20!white](2.7,-1.196)--(2.7,-.732)--(2.433,-.839)--(2.433,-1.304)--cycle;
\filldraw[thin,fill=blue!20!white](.963,-1.196)--(1.364,-1.268)--(1.364,-.804)--(.963,-.732)--cycle;
\filldraw[thin,fill=blue!20!white](-2.7,-.196)--(-2.968,-.304)--(-2.567,-.375)--(-2.3,-.268)--cycle;
\filldraw[thin,fill=blue!20!white](-3.636,-2.196)--(-3.235,-2.268)--(-3.235,-1.804)--(-3.636,-1.732)--cycle;
\filldraw[thin,fill=blue!20!white](-2.3,-1.196)--(-2.3,-.732)--(-2.567,-.839)--(-2.567,-1.304)--cycle;
\filldraw[thin,fill=blue!20!white](-4.037,-1.196)--(-3.636,-1.268)--(-3.636,-.804)--(-4.037,-.732)--cycle;
\filldraw[thin,fill=blue!20!white](.668,-10.196)--(.401,-10.304)--(.802,-10.375)--(1.069,-10.268)--cycle;
\filldraw[thin,fill=blue!20!white](-2.567,-2.232)--(-2.567,-1.768)--(-2.834,-1.875)--(-2.834,-2.339)--cycle;
\filldraw[thin,fill=blue!20!white](2.7,-.732)--(2.7,-.268)--(2.433,-.375)--(2.433,-.839)--cycle;
\filldraw[thin,fill=blue!20!white](2.433,-2.232)--(2.433,-1.768)--(2.166,-1.875)--(2.166,-2.339)--cycle;
\filldraw[thin,fill=blue!20!white](-1.336,-5.232)--(-.935,-5.304)--(-.935,-4.839)--(-1.336,-4.768)--cycle;
\filldraw[thin,fill=blue!20!white](-2.3,-.732)--(-2.3,-.268)--(-2.567,-.375)--(-2.567,-.839)--cycle;
\filldraw[thin,fill=blue!20!white](-3.636,-1.732)--(-3.235,-1.804)--(-3.235,-1.339)--(-3.636,-1.268)--cycle;
\filldraw[thin,fill=blue!20!white](-3.369,-.232)--(-3.636,-.339)--(-3.235,-.411)--(-2.968,-.304)--cycle;
\filldraw[thin,fill=blue!20!white](.963,-.732)--(1.364,-.804)--(1.364,-.339)--(.963,-.268)--cycle;
\filldraw[thin,fill=blue!20!white](1.364,-1.732)--(1.765,-1.804)--(1.765,-1.339)--(1.364,-1.268)--cycle;
\filldraw[thin,fill=blue!20!white](1.631,-.232)--(1.364,-.339)--(1.765,-.411)--(2.032,-.304)--cycle;
\filldraw[thin,fill=blue!20!white](1.47,-5.732)--(1.47,-5.268)--(1.203,-5.375)--(1.203,-5.839)--cycle;
\filldraw[thin,fill=blue!20!white](-.935,-6.232)--(-.535,-6.304)--(-.535,-5.839)--(-.935,-5.768)--cycle;
\filldraw[thin,fill=blue!20!white](0,-4.232)--(-.267,-4.339)--(.134,-4.411)--(.401,-4.304)--cycle;
\filldraw[thin,fill=blue!20!white](0,-8.232)--(-.267,-8.339)--(.134,-8.411)--(.401,-8.304)--cycle;
\filldraw[thin,fill=blue!20!white](0,-10.232)--(-.267,-10.339)--(.134,-10.411)--(.401,-10.304)--cycle;
\filldraw[thin,fill=blue!20!white](-4.037,-.732)--(-3.636,-.804)--(-3.636,-.339)--(-4.037,-.268)--cycle;
\filldraw[thin,fill=blue!20!white](-1.336,-8.768)--(-.935,-8.839)--(-.935,-8.375)--(-1.336,-8.304)--cycle;
\filldraw[thin,fill=blue!20!white](-1.336,-10.768)--(-.935,-10.839)--(-.935,-10.375)--(-1.336,-10.304)--cycle;
\filldraw[thin,fill=blue!20!white](1.47,-5.268)--(1.47,-4.804)--(1.203,-4.911)--(1.203,-5.375)--cycle;
\filldraw[thin,fill=blue!20!white](-3.636,-1.268)--(-3.235,-1.339)--(-3.235,-.875)--(-3.636,-.804)--cycle;
\filldraw[thin,fill=blue!20!white](-2.567,-1.768)--(-2.567,-1.304)--(-2.834,-1.411)--(-2.834,-1.875)--cycle;
\filldraw[thin,fill=blue!20!white](-3.235,-2.268)--(-2.834,-2.339)--(-2.834,-1.875)--(-3.235,-1.804)--cycle;
\filldraw[thin,fill=blue!20!white](1.364,-1.268)--(1.765,-1.339)--(1.765,-.875)--(1.364,-.804)--cycle;
\filldraw[thin,fill=blue!20!white](2.433,-1.768)--(2.433,-1.304)--(2.166,-1.411)--(2.166,-1.875)--cycle;
\filldraw[thin,fill=blue!20!white](1.765,-2.268)--(2.166,-2.339)--(2.166,-1.875)--(1.765,-1.804)--cycle;
\filldraw[thin,fill=blue!20!white](-1.336,-4.768)--(-.935,-4.839)--(-.935,-4.375)--(-1.336,-4.304)--cycle;
\filldraw[thin,fill=blue!20!white](-.935,-5.768)--(-.535,-5.839)--(-.535,-5.375)--(-.935,-5.304)--cycle;
\filldraw[thin,fill=blue!20!white](-.668,-4.268)--(-.935,-4.375)--(-.535,-4.446)--(-.267,-4.339)--cycle;
\filldraw[thin,fill=blue!20!white](1.069,-4.268)--(.802,-4.375)--(1.203,-4.446)--(1.47,-4.339)--cycle;
\filldraw[thin,fill=blue!20!white](-.668,-8.268)--(-.935,-8.375)--(-.535,-8.446)--(-.267,-8.339)--cycle;
\filldraw[thin,fill=blue!20!white](1.069,-8.268)--(.802,-8.375)--(1.203,-8.446)--(1.47,-8.339)--cycle;
\filldraw[thin,fill=blue!20!white](-.668,-10.268)--(-.935,-10.375)--(-.535,-10.446)--(-.267,-10.339)--cycle;
\filldraw[thin,fill=blue!20!white](1.069,-10.268)--(.802,-10.375)--(1.203,-10.446)--(1.47,-10.339)--cycle;
\filldraw[thin,fill=blue!20!white](-2.567,-1.304)--(-2.567,-.839)--(-2.834,-.946)--(-2.834,-1.411)--cycle;
\filldraw[thin,fill=blue!20!white](2.433,-1.304)--(2.433,-.839)--(2.166,-.946)--(2.166,-1.411)--cycle;
\filldraw[thin,fill=blue!20!white](.401,-10.304)--(.134,-10.411)--(.535,-10.482)--(.802,-10.375)--cycle;
\filldraw[thin,fill=blue!20!white](1.47,-10.804)--(1.47,-10.339)--(1.203,-10.446)--(1.203,-10.911)--cycle;
\filldraw[thin,fill=blue!20!white](.401,-8.304)--(.134,-8.411)--(.535,-8.482)--(.802,-8.375)--cycle;
\filldraw[thin,fill=blue!20!white](1.47,-8.804)--(1.47,-8.339)--(1.203,-8.446)--(1.203,-8.911)--cycle;
\filldraw[thin,fill=blue!20!white](.401,-4.304)--(.134,-4.411)--(.535,-4.482)--(.802,-4.375)--cycle;
\filldraw[thin,fill=blue!20!white](1.203,-6.304)--(1.203,-5.839)--(.935,-5.946)--(.935,-6.411)--cycle;
\filldraw[thin,fill=blue!20!white](-.535,-6.304)--(-.134,-6.375)--(-.134,-5.911)--(-.535,-5.839)--cycle;
\filldraw[thin,fill=blue!20!white](-.935,-5.304)--(-.535,-5.375)--(-.535,-4.911)--(-.935,-4.839)--cycle;
\filldraw[thin,fill=blue!20!white](1.47,-4.804)--(1.47,-4.339)--(1.203,-4.446)--(1.203,-4.911)--cycle;
\filldraw[thin,fill=blue!20!white](2.032,-.304)--(1.765,-.411)--(2.166,-.482)--(2.433,-.375)--cycle;
\filldraw[thin,fill=blue!20!white](1.765,-1.804)--(2.166,-1.875)--(2.166,-1.411)--(1.765,-1.339)--cycle;
\filldraw[thin,fill=blue!20!white](1.364,-.804)--(1.765,-.875)--(1.765,-.411)--(1.364,-.339)--cycle;
\filldraw[thin,fill=blue!20!white](-2.968,-.304)--(-3.235,-.411)--(-2.834,-.482)--(-2.567,-.375)--cycle;
\filldraw[thin,fill=blue!20!white](-3.235,-1.804)--(-2.834,-1.875)--(-2.834,-1.411)--(-3.235,-1.339)--cycle;
\filldraw[thin,fill=blue!20!white](-3.636,-.804)--(-3.235,-.875)--(-3.235,-.411)--(-3.636,-.339)--cycle;
\filldraw[thin,fill=blue!20!white](2.433,-.839)--(2.433,-.375)--(2.166,-.482)--(2.166,-.946)--cycle;
\filldraw[thin,fill=blue!20!white](1.765,-1.339)--(2.166,-1.411)--(2.166,-.946)--(1.765,-.875)--cycle;
\filldraw[thin,fill=blue!20!white](-3.235,-1.339)--(-2.834,-1.411)--(-2.834,-.946)--(-3.235,-.875)--cycle;
\filldraw[thin,fill=blue!20!white](-2.567,-.839)--(-2.567,-.375)--(-2.834,-.482)--(-2.834,-.946)--cycle;
\filldraw[thin,fill=blue!20!white](-.935,-4.839)--(-.535,-4.911)--(-.535,-4.446)--(-.935,-4.375)--cycle;
\filldraw[thin,fill=blue!20!white](-.535,-5.839)--(-.134,-5.911)--(-.134,-5.446)--(-.535,-5.375)--cycle;
\filldraw[thin,fill=blue!20!white](1.203,-5.839)--(1.203,-5.375)--(.935,-5.482)--(.935,-5.946)--cycle;
\filldraw[thin,fill=blue!20!white](-.267,-4.339)--(-.535,-4.446)--(-.134,-4.518)--(.134,-4.411)--cycle;
\filldraw[thin,fill=blue!20!white](-.935,-8.839)--(-.535,-8.911)--(-.535,-8.446)--(-.935,-8.375)--cycle;
\filldraw[thin,fill=blue!20!white](-.267,-8.339)--(-.535,-8.446)--(-.134,-8.518)--(.134,-8.411)--cycle;
\filldraw[thin,fill=blue!20!white](-.935,-10.839)--(-.535,-10.911)--(-.535,-10.446)--(-.935,-10.375)--cycle;
\filldraw[thin,fill=blue!20!white](-.267,-10.339)--(-.535,-10.446)--(-.134,-10.518)--(.134,-10.411)--cycle;
\filldraw[thin,fill=blue!20!white](-.134,-6.375)--(.267,-6.446)--(.267,-5.982)--(-.134,-5.911)--cycle;
\filldraw[thin,fill=blue!20!white](-3.235,-.875)--(-2.834,-.946)--(-2.834,-.482)--(-3.235,-.411)--cycle;
\filldraw[thin,fill=blue!20!white](1.765,-.875)--(2.166,-.946)--(2.166,-.482)--(1.765,-.411)--cycle;
\filldraw[thin,fill=blue!20!white](-.535,-5.375)--(-.134,-5.446)--(-.134,-4.982)--(-.535,-4.911)--cycle;
\filldraw[thin,fill=blue!20!white](1.203,-5.375)--(1.203,-4.911)--(.935,-5.018)--(.935,-5.482)--cycle;
\filldraw[thin,fill=blue!20!white](.802,-4.375)--(.535,-4.482)--(.935,-4.554)--(1.203,-4.446)--cycle;
\filldraw[thin,fill=blue!20!white](.802,-8.375)--(.535,-8.482)--(.935,-8.554)--(1.203,-8.446)--cycle;
\filldraw[thin,fill=blue!20!white](.802,-10.375)--(.535,-10.482)--(.935,-10.554)--(1.203,-10.446)--cycle;
\filldraw[thin,fill=blue!20!white](1.203,-4.911)--(1.203,-4.446)--(.935,-4.554)--(.935,-5.018)--cycle;
\filldraw[thin,fill=blue!20!white](-.134,-5.911)--(.267,-5.982)--(.267,-5.518)--(-.134,-5.446)--cycle;
\filldraw[thin,fill=blue!20!white](.935,-6.411)--(.935,-5.946)--(.668,-6.054)--(.668,-6.518)--cycle;
\filldraw[thin,fill=blue!20!white](.134,-4.411)--(-.134,-4.518)--(.267,-4.589)--(.535,-4.482)--cycle;
\filldraw[thin,fill=blue!20!white](-.535,-8.911)--(-.134,-8.982)--(-.134,-8.518)--(-.535,-8.446)--cycle;
\filldraw[thin,fill=blue!20!white](1.203,-8.911)--(1.203,-8.446)--(.935,-8.554)--(.935,-9.018)--cycle;
\filldraw[thin,fill=blue!20!white](.134,-8.411)--(-.134,-8.518)--(.267,-8.589)--(.535,-8.482)--cycle;
\filldraw[thin,fill=blue!20!white](-.535,-10.911)--(-.134,-10.982)--(-.134,-10.518)--(-.535,-10.446)--cycle;
\filldraw[thin,fill=blue!20!white](1.203,-10.911)--(1.203,-10.446)--(.935,-10.554)--(.935,-11.018)--cycle;
\filldraw[thin,fill=blue!20!white](.134,-10.411)--(-.134,-10.518)--(.267,-10.589)--(.535,-10.482)--cycle;
\filldraw[thin,fill=blue!20!white](-.535,-4.911)--(-.134,-4.982)--(-.134,-4.518)--(-.535,-4.446)--cycle;
\filldraw[thin,fill=blue!20!white](.267,-6.446)--(.668,-6.518)--(.668,-6.054)--(.267,-5.982)--cycle;
\filldraw[thin,fill=blue!20!white](-.134,-5.446)--(.267,-5.518)--(.267,-5.054)--(-.134,-4.982)--cycle;
\filldraw[thin,fill=blue!20!white](.935,-5.946)--(.935,-5.482)--(.668,-5.589)--(.668,-6.054)--cycle;
\filldraw[thin,fill=blue!20!white](-.134,-4.982)--(.267,-5.054)--(.267,-4.589)--(-.134,-4.518)--cycle;
\filldraw[thin,fill=blue!20!white](.935,-5.482)--(.935,-5.018)--(.668,-5.125)--(.668,-5.589)--cycle;
\filldraw[thin,fill=blue!20!white](-.134,-8.982)--(.267,-9.054)--(.267,-8.589)--(-.134,-8.518)--cycle;
\filldraw[thin,fill=blue!20!white](-.134,-10.982)--(.267,-11.054)--(.267,-10.589)--(-.134,-10.518)--cycle;
\filldraw[thin,fill=blue!20!white](.535,-8.482)--(.267,-8.589)--(.668,-8.661)--(.935,-8.554)--cycle;
\filldraw[thin,fill=blue!20!white](.535,-4.482)--(.267,-4.589)--(.668,-4.661)--(.935,-4.554)--cycle;
\filldraw[thin,fill=blue!20!white](.267,-5.982)--(.668,-6.054)--(.668,-5.589)--(.267,-5.518)--cycle;
\filldraw[thin,fill=blue!20!white](.535,-10.482)--(.267,-10.589)--(.668,-10.661)--(.935,-10.554)--cycle;
\filldraw[thin,fill=blue!20!white](.935,-9.018)--(.935,-8.554)--(.668,-8.661)--(.668,-9.125)--cycle;
\filldraw[thin,fill=blue!20!white](.935,-11.018)--(.935,-10.554)--(.668,-10.661)--(.668,-11.125)--cycle;
\filldraw[thin,fill=blue!20!white](.267,-5.518)--(.668,-5.589)--(.668,-5.125)--(.267,-5.054)--cycle;
\filldraw[thin,fill=blue!20!white](.935,-5.018)--(.935,-4.554)--(.668,-4.661)--(.668,-5.125)--cycle;
\filldraw[thin,fill=blue!20!white](.267,-11.054)--(.668,-11.125)--(.668,-10.661)--(.267,-10.589)--cycle;
\filldraw[thin,fill=blue!20!white](.267,-9.054)--(.668,-9.125)--(.668,-8.661)--(.267,-8.589)--cycle;
\filldraw[thin,fill=blue!20!white](.267,-5.054)--(.668,-5.125)--(.668,-4.661)--(.267,-4.589)--cycle;
\path[-latex] (-4.638,-2.25) -- (-3.836,-2.393) node[sloped,pos=0.6,below] {$\mathbf{na}$};\path[-latex] (-.87,-1.529) -- (-.87,-.786) node[sloped,pos=0.6,above] {$\mathbf{nsrc}$};\path[-latex] (-2.634,-2.839) -- (-2.099,-2.625) node[sloped,pos=1.1,above] {$\mathbf{ntime}$};\path[-latex] (.362,-2.25) -- (1.164,-2.393) node[sloped,pos=0.6,below] {$\mathbf{na}$};\path[-latex] (4.13,-1.529) -- (4.13,-.786) node[sloped,pos=0.6,above] {$\mathbf{nsrc}$};\path[-latex] (2.366,-2.839) -- (2.901,-2.625) node[sloped,pos=1.1,above] {$\mathbf{ntime}$};\path[-latex] (-2.138,-6.018) -- (-.535,-6.304) node[sloped,pos=0.6,below] {$\mathbf{nbl}$};\path[-latex] (2.673,-5.714) -- (2.673,-4.971) node[sloped,pos=0.6,above] {$\mathbf{nsrc}$};\path[-latex] (1.069,-7.007) -- (1.604,-6.793) node[sloped,pos=1.1,above] {$\mathbf{ntime}$};\node at (-5.5,.5) {$A_{ps}=E_{ps}K_{ps}$};\node at (4,.5) {$A_{qs}=E_{qs}K_{qs}$};\node at (-5.25,-5) {$X_{pqs}=A_{ps}B_{s}A_{qs}$};\node at (-5,-7.5) {$X_{pq}=\sum_{s=0}^{\textrm{nsrc}} X_{pqs}$};\node at (-5,-10) {$D_{pq}$};\node at (-4,-12.25) {$\chi^2=\sum w_{pq}\left( X_{pq} - D_{pq}\right)^2$};\node at (3,-12.25) {$-2\ln\BLikelihoodS+\mathrm{const.}$};\draw[thick,postaction={decorate,decoration={text along path,raise=2mm,text align=right,text={EK kernel}}}] (-7,1) -- (-7,-2);\draw[thick,postaction={decorate,decoration={text along path,raise=2mm,text align=right,text={B Sum kernel}}}] (-7,-3) -- (-7,-10.5);\draw[thick,postaction={decorate,decoration={text along path,raise=2mm,text align=right,text={Reduce}}}] (-7,-11) -- (-7,-13);\end{tikzpicture}% End sketch output
\end{center}
\caption{Workflow for computing the BIRO RIME. Firstly,
  per-antenna $A_{ps}$ values are computed in the EK kernel
  and combined to form  per-baseline $X_{pqs}$ source-coherency terms
  in the B Sum kernel.
  The source  coherencies are summed to form model visibilities $X_{pq}$, which
  are subtracted from the observed visibilities $D_{pq}$, squared, and
  multiplied by weight $w_{pq}$. Finally this result is reduced/summed
  by a reduction kernel, to form a $\chi^2$ value from which
  the likelihood follows trivially using equation~\ref{eqn:lhood}.}
\label{fig:workflow}
\end{figure}

\section{Architecture}

In this section, we discuss Montblanc's RIME architecture
and implementation.

As discussed in the previous section, neither
\textsc{MeqTrees} or \textsc{OSKAR} support
evaluating the entire RIME on GPU, nor do they
provide the facility to calculate a $\chi^2$ value.
As BIRO makes many RIME evaluations, it is
important to support this iterative approach
to computation of the $\chi^2$. For this reason,
Montblanc computes the entire RIME
and $\chi^2$ on the GPU (Figure \ref{fig:biro_workflow}).
The sky model, telescope configuration and observed
visibilities are transferred to the GPU
on each iteration and a single,
floating point $\chi^2$ is
transferred off.
Data transfer to the GPU is
overlapped by GPU kernel execution
for a sufficient number of sources
(See Section \ref{sec:rime_scaling}).

\subsection{Data and ordering}

\noindent
The input for the RIME (Table~\ref{tbl:rime_input}) is a series of 1D,
2D and 3D arrays, of which some are obtained from a CASA Measurement
Set (MS) file \citep{McMullin2007,CASA2014}. A workflow
(Figure~\ref{fig:workflow}) is applied to solve the RIME and compute
the $\chi^2$ value (the second normalization constant term is a
function of the weight vector only and trivially calculated on the CPU
via equation~\ref{eqn:lhood}).

Firstly, the arrays are combined to create per-antenna values. The
strategy here is to perform computationally expensive operations
per-antenna, rather than per-baseline which is quadratically-greater
in number.

The per-antenna terms are combined to create a 4D source-coherency
array, followed by reduction to a 3D model visibility array. The model
visibilities are subtracted from the data, squared, and multiplied by
the weight vector, then summed to produced a single scalar $\chi^2$
value. Each of the 4D arrays' values and 3D reductions can be computed
independently.

In general, we order our GPU arrays by $\mathtt{ntime \times nbl
  \times nsrc \times nchan}$ where {\tt ntime} and {\tt nchan} are the
slowest and most rapidly changing dimensions respectively. {\tt ntime}
and {\tt nbl} are chosen as the first and second dimensions since the
ordering of data in a CASA measurement set (MS) MAIN table is usually
\footnote{$\mathtt{nbl \times ntime}$ CASA orderings are possible but
  Montblanc does not support them.} $\mathtt{ntime \times nbl}$. Using
the same ordering as the MS avoids overly complex transpose operations
and allows for efficient streaming from disk. In support of this, our CUDA
kernels are 3D kernels with grid sizes of $\mathtt{ntime \times na
  \times nchan}$ and $\mathtt{ntime \times nbl \times nchan}$.

Each kernel loops over the source dimension. This is useful because
the computation for point and Gaussian sources is different,
making parallelisation over the source dimension unwieldy. New source
types can be supported by adding another source loop to the kernel.
Looping also increases the amount of work performed by each kernel thread,
amortising the cost of kernel launch and clean-up \citep{CUDA}.
A 3D kernel configuration facilitates loading of per-timestep, -antenna
and -channel data into shared memory at the start of the kernel.
At each source-loop iteration, these data are combined with source data
to form a tensor product. As each thread is accessing the same
or similar source data simultaneously, this data will be retained in
L1/L2 cache, increasing the opportunity for cache hits to occur.

Furthermore, the channel dimension is the easiest to parallelise since
it is a 1D array and the same operation can be computed
simultaneously for different wavelengths. This exploits CUDA's SIMD
architecture: each value is independent, leading to coalesced reads and writes
to global memory.

\begin{table}
\begin{center}
\begin{tabular}{|l|c|c|c|}
\hline
Kernel & Registers & Threads per Block & Occupancy \\
\hline
EK float & 31 & 512 & 87\% \\
EK double & 44 & 256 & 59\% \\
\hline
B Sum float & 46 & 256 & 60\% \\
B Sum double & 63 & 128 & 50\% \\
\hline
\end{tabular}
\caption{Kernel Configuration. This table shows the
number of registers, threads per block and the
occupancy of each kernel, as reported by the
NVIDIA Profiler \citep{CUDA_PROFILER}.}
\label{tbl:kernel_config}
\end{center}
\end{table}

\begin{algorithm}
\begin{algorithmic}
\Procedure{EKKernel}{time $\mathtt{t}$,antenna $\mathtt{p}$,channel $\mathtt{\lambda}$}
\State Read {\itshape uvw} coordinates and wavelengths from RAM into shared memory.
\ForAll{point sources $s$}
\State Read {\itshape lm} coordinates into shared memory.
\State Synchronize Threads
\State Compute tensor $\mathtt{K[t,p,s,\lambda]}$
\State Compute tensor $\mathtt{E[t,p,s,\lambda]}$
\State Compute $\mathtt{A[t,p,s,\lambda] = E[t,p,s,\lambda] \times K[t,p,s,\lambda]}$
\State Write $\mathtt{A[t,p,s,\lambda]}$ to DRAM.
\State Synchronize Threads
\EndFor
\ForAll{Gaussian sources $s$}
\State Read {\itshape lm} coordinates into shared memory.
\State Synchronize Threads
\State Compute tensor $\mathtt{K[t,p,s,\lambda]}$
\State Compute tensor $\mathtt{E[t,p,s,\lambda]}$
\State Compute $\mathtt{A[t,p,s,\lambda] = E[t,p,s,\lambda] \times K[t,p,s,\lambda]}$
\State Write $\mathtt{A[t,p,s,\lambda]}$ to DRAM
\State Synchronize Threads
\EndFor
\EndProcedure
\end{algorithmic}
\caption{The EK kernel is parallelised over time, antennas and channel
  dimensions. Each source type is handled as a loop, in which the
  per-antenna value for each source is computed and outputted.
  Extending Montblanc to support $\beta$ profiles would involve
  adding a new loop to the kernel.}
\label{alg:ek_kernel}
\end{algorithm}

\begin{algorithm}
\begin{algorithmic}
\Procedure{BSumKernel}{time $\mathtt{t}$,baseline $\mathtt{pq}$, channel $\mathtt{\lambda}$}
\State $\mathtt{V[t,pq,\lambda] \gets 0}$
\State Read {\itshape uvw} coordinates and wavelengths from DRAM into shared memory.
\ForAll{$s$ in point sources}
\State Read $\mathtt{B[t,s]}$ into shared memory.
\State Synchronize Threads
\State Read $\mathtt{A[t,p,s,\lambda]}$ and $\mathtt{A[t,q,s,\lambda]}$ from DRAM
\State $\mathtt{X[t,pq,\lambda,s] = A[t,p,s,\lambda] \times B[t,s] \times  A[t,p,s,\lambda].H}$
\State $\mathtt{V[t,pq,\lambda] += X[t,pq,\lambda,s]}$
\State Synchronize Threads
\EndFor
\ForAll{$s$ in gaussian sources}
\State Read $\mathtt{B[t,s]}$ into shared memory.
\State Synchronize Threads
\State Read $\mathtt{A[t,p,s,\lambda]}$ and $\mathtt{A[t,q,s,\lambda]}$ from DRAM
\State $\mathtt{X[t,pq,\lambda,s] = A[t,p,s,\lambda] \times B[t,s] \times  A[t,p,s,\lambda].H}$
\State $\mathtt{V[t,pq,\lambda] += X[t,pq,\lambda,s]}$
\State Synchronize Threads
\EndFor
\State Output $\mathtt{V[t,pq,\lambda]}$ to DRAM \Comment Optional
\State Read $\mathtt{w[t,pq,\lambda]}$, $\mathtt{D[t,pq,\lambda]}$ from DRAM
\State Output $\mathtt{w[t,pq,\lambda] \times \left( V[t,pq,\lambda] - D[t,pq,\lambda] \right)^2\equiv\chi^2[t,pq,\lambda]}$ to DRAM
\EndProcedure
\end{algorithmic}
\caption{The B sum kernel is parallelised over time, baseline and
  channel. The visibility is initialised and a loop is employed for
  each source type. Within each loop, the appropriate per-antenna
  values for the baseline are read, and combined with the source
  brightness matrix to form a source coherency. These are added to the
  visibilities. Once the model visibilities have been calculated, they
  are differenced with the observed visibilities to calculate a
  $\chi^2_{pq}$ term for each visibility.}
\label{alg:bsum_kernel}
\end{algorithm}

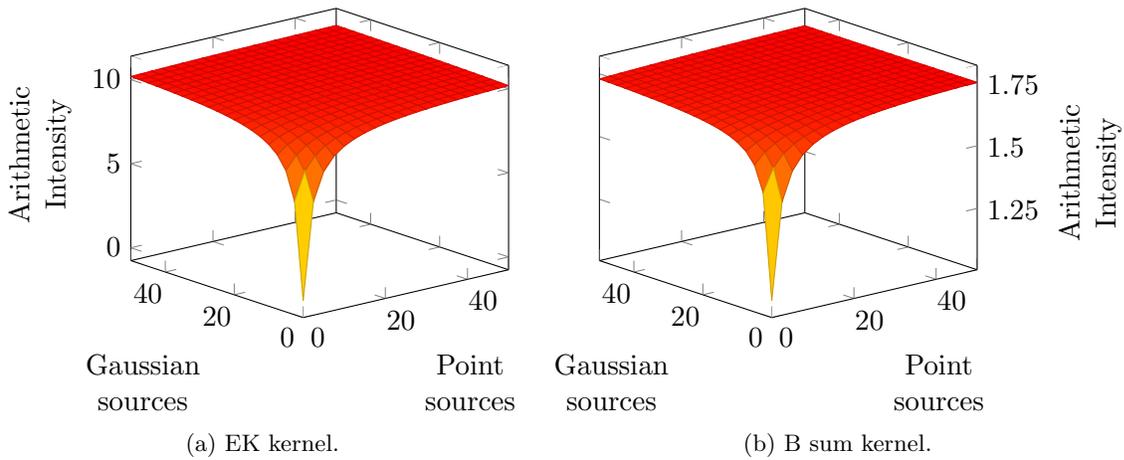
\begin{figure}
\subfloat[EK kernel.]
{
\begin{tikzpicture}
\begin{axis}[
    width=0.4\textwidth,
    view={320}{20},
    samples=21,
    domain=0:50,
    y domain=0:50,
    xlabel style={align=center},
    ylabel style={align=center},
    zlabel style={align=center},
    xlabel={Point\\sources},
    ylabel={Gaussian\\sources},
    zlabel={Arithmetic\\Intensity}
]
\addplot3[surf] {(6 + 127*(x+y))/(4*(6 + 3*(x+y)))};
\end{axis}
\end{tikzpicture}
\label{fig:ek_intensity}
}
\subfloat[B sum kernel.]
{
\begin{tikzpicture}
\begin{axis}[
    width=0.4\textwidth,
    view={320}{20},
    samples=21,
    domain=0:50,
    y domain=0:50,
    ztick pos=right,
    ztick={1.0,1.25,1.5,1.75,2.0},
    xlabel style={align=center},
    ylabel style={align=center},
    zlabel style={align=center},
    xlabel={Point\\sources},
    ylabel={Gaussian\\sources},
    zlabel={Arithmetic\\Intensity}
]
\addplot3[surf] {(73 + 57*x + 77*y)/(4*(17 + 8*x + 11*y))};
\end{axis}
\end{tikzpicture}
\label{fig:bsum_intensity}
}
\caption{Arithmetic Intensities of the EK and B Sum kernel, for a
  given number of point and Gaussian sources.}
\label{fig:intensities}
\end{figure}

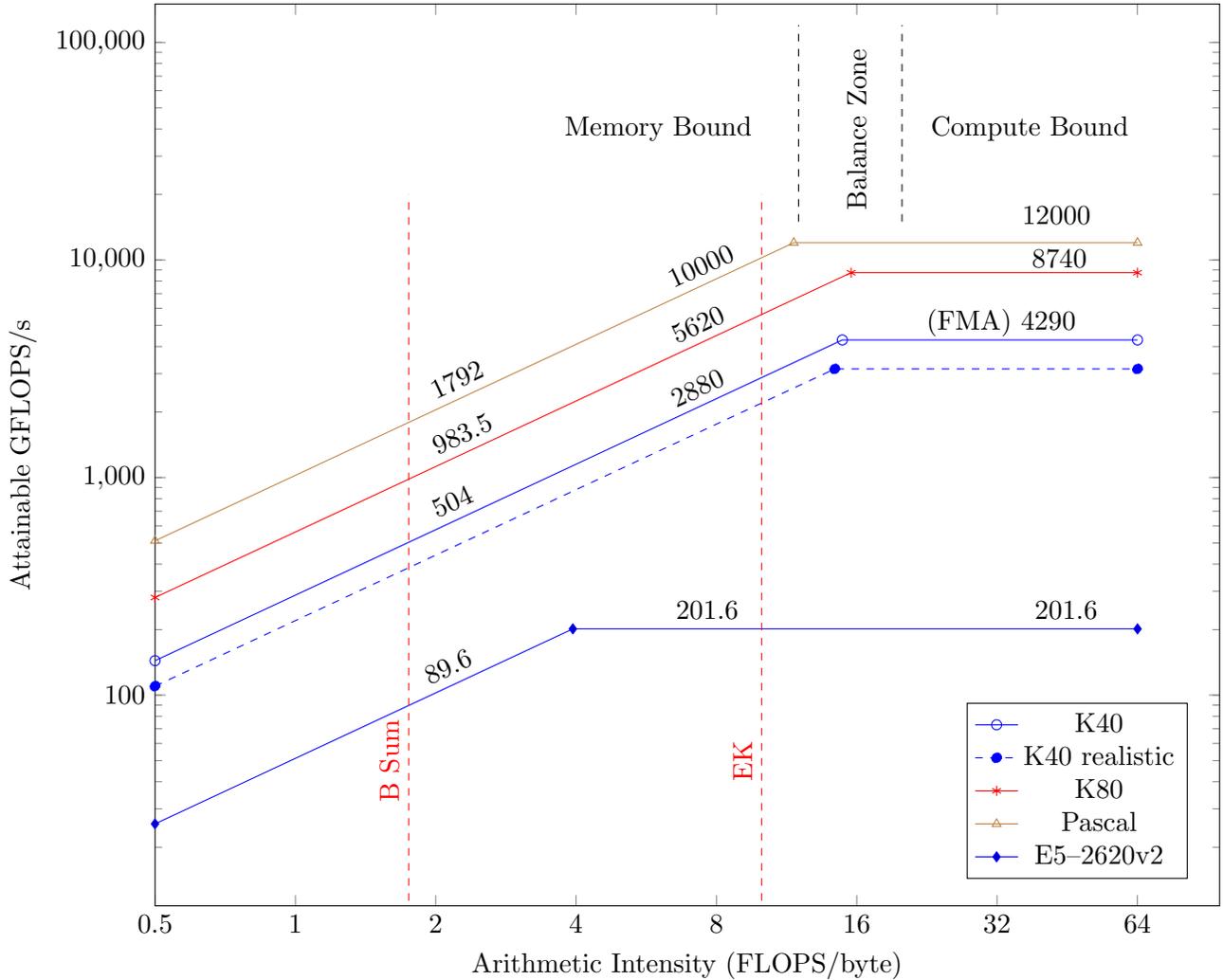
\begin{figure}
\begin{tikzpicture}
\tikzset{addflops/.style n args={3}{
    decorate,
    decoration={
        text along path,
        text align={left indent={#1\dimexpr\pgfdecoratedpathlength\relax}},
        raise={#2},
        text={#3}
    }
}}
\begin{axis}[
    width=1.0\textwidth,
    xmode = log,
    ymode = log,
    log basis x=2,
    log ticks with fixed point,
    xmin=0.5, xmax=96,
    ymin=0, ymax=150000,
    %x filter/.code=\pgfmathparse{#1 * 2},
    %xtick={0, 0.25, 0.5, 1, 2, 4, 8, 16},
    %xticklabels={0, $\frac{1}{4}$, $\frac{1}{2}$, 1, 2, 4, 8, 16},
    xlabel=Arithmetic Intensity (FLOPS/byte),
    ylabel={Attainable GFLOPS/s},
    legend entries={K40, K40 realistic, K80, Pascal, E5--2620v2},
    legend pos=south east]
% NVIDIA K40 (single)
% 4290 GFLOPS/s, 288 GB/s
\addplot[color=blue,mark=o]
coordinates { %K40
    (0.5, 0.5*288)
    (4290/288, 4290)
    (64, 4290)
}
[postaction={addflops={0.3}{4pt}{504}}]
[postaction={addflops={0.55}{4pt}{2880}}]
[postaction={addflops={0.8}{4pt}{(FMA) 4290}}];

% More realistic K40 plot
\addplot[blue, mark=*, dashed, sharp plot, update limits=false]
coordinates {
    (0.5, 0.5*220)
    (3156/220, 3156)
    (64, 3156)
};

% NVIDIA K80
% 8740 GFLOPS/s, 562 GB/s
\addplot[red, mark=asterisk]
coordinates { %K80
    (0.5, 0.5*562)
    (8740/562, 8740)
    (64, 8740)
}
[postaction={addflops={0.3}{4pt}{983.5}}]
[postaction={addflops={0.55}{4pt}{5620}}]
[postaction={addflops={0.9}{2pt}{8740}}];

% NVIDIA Pascal
% 12000 GFLOPS/s, 1000 GB/s
\addplot[brown, mark=triangle]
coordinates { % Pascal
    (0.5, 0.5*1024)
    (12000/1024, 12000)
    (64, 12000)
}
[postaction={addflops={0.3}{4pt}{1792}}]
[postaction={addflops={0.55}{4pt}{10000}}]
[postaction={addflops={0.89}{7.5pt}{12000}}];

% Intel (R) Xeon CPU Et-2620v2 @ 3.00Ghz and max bandwidth of 51.2GB/s
\addplot coordinates {
   (0.5, 0.5*51.2)
   (201.6/51.2, 201.6)
   (64, 201.6)
}
[postaction={addflops={0.3}{4pt}{89.6}}]
[postaction={addflops={0.55}{4pt}{201.6}}]
[postaction={addflops={0.9}{4pt}{201.6}}];

% EK kernel arithmetic intensity line
\addplot[red,dashed,sharp plot,update limits=false]
    coordinates {(10,0.25) (10,20000)}
    node[above,rotate=90] at (axis cs:10,50) {EK};

% B Sum kernel arithmetic intensity line
\addplot[red!100,dashed,sharp plot,update limits=false]
    coordinates {(1.75,0.25) (1.75,20000)}
    node[above,rotate=90] at (axis cs:1.75,50) {B Sum};

% Memory bound line and text
\addplot[black,dashed, sharp plot, update limits=false]
            coordinates {(12,15000) (12,120000)}
            node[left] at (axis cs:10,40000) {Memory Bound};

% Compute bound line and text
\addplot[black,dashed, sharp plot, update limits=false]
            coordinates {(20,15000) (20,120000)}
            node[right] at (axis cs:22,40000) {Compute Bound};

% Balance zone line and text
\addplot[black,dashed, sharp plot, update limits=false]
            coordinates {(0,10000) (128,10000)}
            node[rotate=90] at (axis cs:16,40000) {Balance Zone};

            %node[rotate=90] at (axis cs:16,40000) {Balance Zone};

\end{axis}
\end{tikzpicture}
\caption{A roofline model of multiple GPU architectures, as well
  as that of 2 hexacore Xeon E5--2620 v2 CPUs. This model
  provides a theoretical estimate of an algorithm's
  attainable GFLOPS/s, given it's arithmetic intensity,
  on different computing architectures.
  The numbers shown for these devices are only attainable
  through the use of {\itshape fused multiply adds} (FMA)
  on data in registers.
  Generally, kernels consist of a much wider range of
  instructions and memory types,
  and therefore their peak performance is correspondingly lower.
  For example the dashed blue line illustates a more realistic
  performance pattern.
  Kernels with low arithmetic intensity cannot approach
  a device's maximum performance, or {\itshape roof}, and are classified
  as {\itshape memory bound.} They must wait for data
  in order to exercise the processor, By contrast, high arithmetic
  intensity algorithms do approach the roof by executing many instructions,
  and are classified as {\itshape compute bound}.
  The point at which the arithmetic
  intensity changes from memory to compute bound is the
  {\itshape balance zone}. This occurs at approximately
  14 FLOPS/byte for Kepler devices and 3.93 FLOPS/byte for the Xeons.
  This model suggests that the two kernels are
  memory bound on Keplers, with
  arithmetic intensity of 1.75 and 10 FLOPS/byte respectively.
  However, these kernels are composed of more general
  instructions and in practice, their balance zone would occur at
  lower arithmetic intensity.
  Determing the compute or memory bound nature
  of these kernels is accomplished through {\itshape profiling}.}
\label{fig:roofline}
\end{figure}

\subsection{Kernels for computing the BIRO RIME solution}

\noindent
Calculating the RIME is accomplished through
three CUDA kernels. The number of
registers, threads per block and occupancy for the first two
are shown in Table \ref{tbl:kernel_config}.

(i) The {\bf EK kernel} (Algorithm \ref{alg:ek_kernel}) computes
per-antenna terms, $\mathtt{A[t,p,s,\lambda]}$, using the analytic expressions in
Equations \ref{eqn:k_term} and \ref{eqn:e_term}. It is parallelised
over the {\tt ntime}, {\tt na} and {\tt nchan} dimensions and
loops over the {\tt nsrc} dimension. A complex scalar
$\mathtt{A[t,p,s,\lambda] = E[t,p,s,\lambda] \times K[t,p,s,\lambda]}$
is generated for each source and the resulting $\mathtt{ntime
 \times na \times nsrc \times nchan}$ array is written to the GPU's
global memory for use by the B sum kernel (see below). In practice, this
kernel is inexpensive compared to others and therefore contains the
computationally expensive trigonometric and transcendental functions
of the RIME.

(ii) The {\bf B sum kernel} (Algorithm \ref{alg:bsum_kernel}) combines the
$\mathtt{A[t,p,s,\lambda]}$ terms produced by the EK kernel, along with a source
brightness matrix $\mathtt{B[t,s]}$, to form per-baseline source coherencies
$\mathtt{X[t,pq,\lambda,s]}$. It is parallelised over the {\tt ntime}, {\tt nbl} and
{\tt nchan} dimensions and also loops over the {\tt nsrc}
dimension. Internally, the loop computes source coherencies and sums
them to produce a $\mathtt{ntime \times nbl \times nchan}$ array of
model visibilities $\mathtt{V[t,pq,\lambda]}$. The model visibilities are then
subtracted from the observed visibilities $\mathtt{D[t,pq,\lambda]}$, squared and
multiplied with the weight vector $\mathtt{w[t,pq,\lambda]}$ to produce a $\mathtt{ntime
  \times nbl \times nchan}$ array of $\chi^2$ terms. This kernel is
the most expensive since the space of values it must process is of size
$\mathtt{ntime \times nbl \times nsrc \times nchan}$.
It is not entirely necessary to write the visibilities to the GPU's
global memory at this point, but this functionality allows Montblanc
to act as a simulator.

(iii) The final {\bf reduction kernel} is a standard reduction
\citep{Harris2005,Harris2007} of the $\chi^2_{pq}$ terms to produce a
single $\chi^2$ value from which the likelihood $\BLikelihoodS$ can be
trivially calculated on the CPU using equation~\ref{eqn:lhood}.

\subsection{Kernel Analysis}

 \begin{table}
\begin{center}
\begin{tabular}{|l|r|r|}
\hline
Operation & float & double  \\
\hline
\multicolumn{3}{|c|}{EK Kernel} \\
\hline
Instruction & 68\% & 75\% \\
Memory & 15\% & 35\% \\
\hline
\multicolumn{3}{|c|}{B Sum Kernel} \\
\hline
Instruction & 65\% & 62\% \\
Memory & 35\% & 45\% \\
\hline
\end{tabular}
\end{center}
\caption{The NVIDIA profiler's \citep{CUDA_PROFILER} estimate
of the percentage of time spent executing instructions and memory loads.
The profiler considers the EK kernel to be compute bound, while it considers
the B sum kernel to be balanced: neither compute nor memory bound.}
\label{tbl:compute_or_memory_bound}
\end{table}

\begin{table}
\begin{center}
\begin{tabular}{|l|r|r|r|r|}
\hline
Operations & bandwidth   & \% of  & bandwidth      & \% of  \\
                 & GB/s (float) &  peak & GB/s (double)  &  peak \\
\hline
Shared Loads & 398.424  & - & 334.269  & - \\
Shared Stores & 147.775  & - & 139.231  & - \\
Global Loads & 132.189  & - & 205.546 & - \\
Global Stores & 2.758  & -& 4.298  & - \\
L1/Shared Total & 681.147  & $\approx$ 30\%  & 683.344  & $\approx$ 20\% \\
\hline
L2 Cache Total & 134.987  & $\approx$ 30\% & 209.844  & $\approx$ 40\% \\
\hline
Device Memory Total & 8.108  & $\approx$ 10\% & 12.619  & $\approx$ 10\% \\
\hline
\end{tabular}
\end{center}
\caption{Memory Performance of the B sum kernel as reported
by the NVIDIA Profiler \citep{CUDA_PROFILER}.
Global Loads/Stores are reads from DRAM, or global memory,
while Shared Loads/Stores are reads from shared memory.
These four figures are totalled under L1/Shared.
Global operations are grouped with L1 cache
because they go through the L1 cache in Fermi devices.
This is not the case for Kepler devices.
Global operations do go through the L2 cache for
both architectures and the sum of
Global Loads/Stores equals the L2 Cache total.
The utilisation of L1 and L2 cache is high relative
to DRAM, or Device Memory.
This indicates many cache hits,
resulting in the avoidance of
expensive DRAM requests.}
\label{tbl:memory_performance}
\end{table}

We have analysed the arithmetic intensity
of the EK and B Sum kernels, estimating
the number of FLOPS for the {\tt sinf} (14), {\tt cosf} (14), {\tt
  sincosf} (14), {\tt expf} (11) and {\tt powf} (54) functions based
on the {\tt \_\_internal\_accurate} functions in CUDA's {\tt
  math\_functions.h} header file. Note that many of these instructions
  are composed of FMA's to approximate Taylor series for example.
As the kernels loops through sources, the FLOPs and byte reads
are dependent on the number of point (P) and Gaussian (G) sources.
We analysed each kernel, line by line, to determine
these counts and found that they are governed by
the following expressions:

\begin{align}
\textrm{EK arithmetic intensity} & = \frac{6 + 127(P+G)}{4(6 + 3(P+G))} \label{eqn:ek_ai} \\
\textrm{B sum arithmetic intensity} & = \frac{73 + 57P + 77G}{4(17 + 8P + 11G) } \label{eqn:bsum_ai}
\end{align}

\noindent
Figure \ref{fig:intensities} shows how this intensity varies for a
given number of sources. It can be seen that the arithmetic intensity
of the EK kernel is $\approx 10$, while that of the
B sum kernel is $\approx 1.75$ and these numbers hold for
varying numbers of point and Gaussians sources. In practice, these values
will be higher since each block and thread accesses the same
values for each source, resulting in L1 and L2 cache hits.

It is useful to plot the arithmetic intensity of these kernels
on a {\itshape Roofline Model} (Figure \ref{fig:roofline}).
This model relates the theoretical peak performance
of an algorithm in GFLOPS/s for different compute architectures,
given it's arithmetic intensity. The K40 and dual hexacore Xeons have
a theoretical maximum, or roof, of 4290 and 201.6 GFLOPS/s
respectively, and a maximum memory bandwidth
of 288 GB/s and 51.2 GB/s respectively.
The disadvantage of the dual Xeon architecture,
compared to the GPUs is clearly visible in the model: A
Xeon Phi 5110P with memory bandwidth of 320 GB/s and
60 cores would be more competitive.
The trend for CPU and GPU architectures to increase in both
performance and memory bandwidth continues. The
values plotted at the intersection arithmetic intensity and
GFLOPS/s show projected performance.

{\itshape Compute bound} algorithms have high arithmetic
intensity and are ideal in the sense that they are bound
by the device's peak performance.
At 10 FLOPS/byte, the EK kernel approaches this limit.
By contrast,
the B Sum kernel is {\itshape memory bound} at 1.75 FLOPS/byte.
Such algorithms have low arithmetic intensity and, being
more dependent on memory reads, do not fully
exercise a device's compute capabilities.
The point at which an algorithm becomes less dependent
on data and more constrained by instruction execution,
occurs between compute and memory bound kernels
in the {\itshape balance zone}. For example, this zone
occurs at $3.93$ FLOPS/byte for Xeons and $\approx 14$
FLOPS/byte for Kepler devices

In practice, this performance only occurs for FMA instructions
operating on data in registers.
More general algorithms have difficulty fully exercising a
device's compute and memory bandwidth.
This means that the balance zone
actually occurs in areas of lower arithmetic intensity.
To discover the compute or memory bound nature
of the kernels, we used the NVIDIA profiler and recorded the
results in Table \ref{tbl:compute_or_memory_bound}.
The profiler considers the EK kernel to be compute bound,
spending most of it's time executing intructions.
This agrees with the kernel's high arithmetic intensity rating.
However, the B sum kernel spends comparitively more time
on memory loads. The profiler considers it to be
balanced -- neither compute nor memory bound.
This is surprising considering it's
relatively low arithmetic intensity.
To see why this is the case, we must examine how
the K40's {\itshape cache} responds to the kernel
(Table \ref{tbl:memory_performance}).

For loads, the ratio of L1 and L2 cache bandwidth to
device memory, or DRAM bandwidth is high,
$(334.269 + 205.546)/12.619 = 42.7 \times$
in the double precision case.
This behaviour occurs because, parallelised over multiple
timesteps, baselines and frequencies, we loop over sources $s$,
synchronising all threads.
Therefore, all SMX's handle one source
in parallel and consequently, a small portion of the
brightness array is simultaneously available in L2 cache to
all SMX's. e.g. $\mathtt{B[t,6]}$ for point source 6.
Additionally, other data such as {\itshape uvw} coordinates and
{\itshape wavelength} is also stored in shared memory (L1 cache).
Retaining significnat portions of the problem in cache means
that many slower trips to DRAM can be avoided:
the caching capabilities of Kepler have hidden the
memory bound nature of the B sum kernel,
making it appear balanced.

We also investigated using Kepler's L1 read cache to store source
and per-antenna data, but did not find a significant
performance difference with our current approach.

\subsection{Scaling the RIME}
\label{sec:rime_scaling}

\begin{figure}
\begin{center}
\includegraphics[width=.95\textwidth]{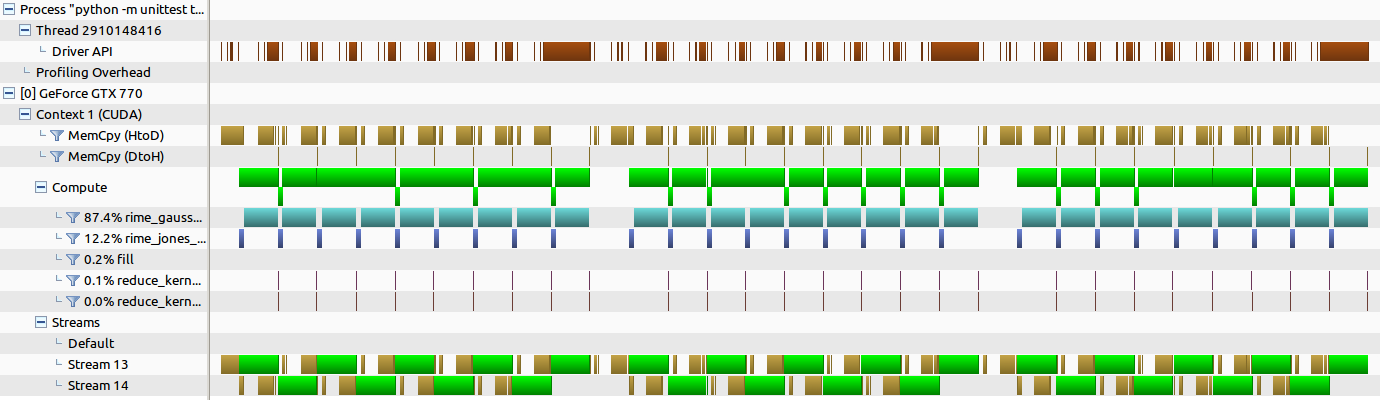}
\end{center}
\caption{Asynchronous Overlap of Data Transfer and Kernel Execution.
This figure shows the computation of three separate solutions to the RIME
for 100 sources. Computation tends to completely overlap data transfer
when the number of sources $\simeq 75$.}
\label{fig:async_overlap}
\end{figure}

\begin{algorithm}
\begin{lstlisting}[language=Python]
import montblanc
# Obtain a solver
with montblanc.get_biro_solver('WSRT.MS',version='v3',
    npsrc=10,ngsrc=10,dtype=numpy.float32) as slvr:

    # Register a UVW array of floating point values
    slvr.register_array(name='uvw',
        shape=(3,'ntime','nbl'), dtype='ft')
    # Transfer some random UVW coordinates to the GPU
    slvr.transfer_uvw(np.random.random(
        shape=slvr.uvw_shape, dtype=slvr.uvw_dtype))
    # Register a property
    slvr.register_property(name='ref_wave',
        dtype='ft',default=1.4e9,value=1.4e9)
    # Set the property
    slvr.set_ref_wave(0.2)

    slvr.solve()     # Solve the BIRO RIME
    print slvr.X2    # Print the Chi-squared value
\end{lstlisting}
\caption{Sample Montblanc code. A solver is obtained that loads the
  interferometer configuration from the WSRT.MS measurement set, and
  specifies 10 point and gaussian sources and single floating-point
  precision. A \textit{uvw} array is registered on the solver and
  random (by way of illustration) \textit{uvw} coordinates transferred
  to the GPU. A reference wavelength property is also registered on
  the solver. Thereafter, the RIME is solved and the $\chi^2$ value
  is retrieved from the solver object.}
\label{alg:code}
\end{algorithm}

\noindent
If the dimensions involved in the RIME are small enough, the
required data may be small enough to fit into the
memory of a single GPU. In practice, this will not be the case,
considering how these dimensions scale for forthcoming
interferometers: for example, MeerKAT will have 64 antennas, 2016
baselines and 32,000 channels, while the SKA will have $\approx 3600$
antennas, $\approx 6.5$ million baselines and 256,000
channels. Solving the RIME for the SKA instrument may require highly
specialised or new computing architectures. However, it is possible to
subdivide the RIME into parts that can be solved within the
memory budgets on contemporary hardware architectures.

As the values of the RIME terms can be calculated independently,
computation of the RIME can be divided so that the problem fits
within the memory budget of a single GPU, individual
components can be solved on multiple GPUs or even
multiple GPUs on physically separate computers.

Here, the
$\mathtt{ntime \times (nbl|na) \times nsrc \times nchan}$ ordering
chosen for Montblanc's imposed on our arrays is important since
each input, temporary and output array adheres to it
\footnote{Other orderings are certainly possible and the
technique described in this section would apply to them too}.
Additionally, each array is registered in a manner similar to a numpy array,
with an array {\itshape shape} and {\itshape type}.
Montblanc takes this further by allowing strings naming each dimension
to be used in addition to integer shape parameters.
See the {\tt register\_array} method
in Algorithm \ref{alg:code} for example.

Once the problem size is known, string parameters are
replaced with integer dimensions and the amount of memory required
for each array and the total problem is derived. Then, if the problem
is too large to fit into a supplied memory budget, the problem is subdivided
along the time dimension. For example, if a problem of size
$\mathtt{ntime \times nbl \times nsrc \times nchan}$ cannot fit into a
GPU memory, but $\mathtt{4 \times nbl \times nsrc \times nchan}$ can,
then two solvers handling problems of size
$\mathtt{2 \times nbl \times nsrc \times nchan}$ are created which solves for
a $\chi^2$ over 2 timesteps, all baselines, sources and channels. Then, we
iterate over timesteps, transferring data related to those timesteps into
solvers. Each solver solves it's part of the $\chi^2$ and the result
is added together to a $\chi^2$ total on the CPU.
By allocating two solvers, we take
advantage of CUDA's {\itshape asynchronous} architecture: The data
required for one solver can be transferred to the GPU {\itshape asynchronously}
while the other solver is performing computation.
As the input (3D) and output (1 scalar) are
smaller relative to the amount of computation (4D), the BIRO
RIME is again well-suited to GPU implementation. Through
empirical tests we have observed that the latency incurred by
data transfer is fully overlapped by execution
when $\simeq 75$ sources are specified (See Figure \ref{fig:async_overlap}).
Consequently we categorise the RIME as {\itshape Dependent-Streaming}
according to the taxonomy of \cite{Gregg2011}.

At present, subdividing along the time dimension suffices for problem sizes
posed by current telescopes, such as Westerbork and LOFAR. However, newer
telescopes will have more baselines and channels, increasing their sensitivity.
They will therefore be able to detect fainter sources, increasing
the number of sources in BIRO's sky models.

However, the subdivision strategy described above can simply be extended
down the row-major ordering. For example, if it is not possible
to fit all timesteps and baselines into a GPU memory, solvers can be allocated
catering for a problem size of $\mathtt{1 \times 2 \times nsrc \times nchan}$:
each solver now solves one timestep, two baselines, all sources and all channels.
However, increasing subdivision means that the problem size is outstripping
 the parallelism afforded by a single GPU.

Thus, increasing time, channel and source dimensions pose a
computational challenge to solving the RIME. One mitigating factor is the
continually increasing computing power and memory of modern GPUs.
A K80 Kepler features 24GB of RAM over
a K40's 12GB, and the newer NVIDIA Pascal architectures will feature
$\pm$ 12000 GFLOP/s, compared to the K40's 4290 GFLOPS/s.

 The obvious solution is to utilise multiple GPU and
 clustered solutions to the to recover some parallelism.
 Properly considered, this raises interesting questions as to whether
 it is even feasible to store the observed visibilities in a Measurement Set
 file. Instead, it may be necessary to stream these visibilities directly
 off a telescope into the RIME solver. We intend to explore these directions
 in future work.

 One other possible avenue is to take an information theoretic approach and
 examine which RIME inputs actually contribute to the $\chi^2$ value.
 Certain channels, baselines and sources may have negligble impact.
 It remains to be seen whether discarding information negatively impacts the
 Bayesian inference process.

\subsection{Implementation Details}
\label{sec:implementation}

\noindent
Montblanc was developed in Python,
using the numpy \citep{numpy, VanDerWalt2011} and PyCUDA
\citep{Kloeckner2012_PyCUDA} packages, and is publicly available at \\
\url{https://github.com/ska-sa/montblanc} under a GPL2 license.

The architecture is designed to be highly modular and extensible.
A contributor \cite{Rivi2015} has already added Sersic profiles
\cite{sersic63} for example.
The basis for the architecture is a {\itshape Solver} object, on which
the arrays used to solve the RIME are registered.
Registering an array automatically creates numpy arrays on the CPU
and CUDA arrays on the host, as well as transfer functions for moving
data from the CPU arrays to GPU arrays. Scalar properties can also be
registered on the solver. A simple usage pattern is demonstrated in Algorithm
\ref{alg:code}.

Developers are not restricted to using the Montblanc RIME for BIRO.
The Solver executes a pipeline of GPU kernels which can be flexibly
configured to suit a particular use case. Such kernels are represented
as Python string templates, allowing a developer to embed string
constants into the kernel.

Montblanc supports both single- and double-precision kernels. This is
useful since single-precision floating-point accuracy will degrade
once RIME dimensions scale upwards, especially in kernel sums and
reductions. It may be possible to employ Kahan sums to reduce this
\citep{Kahan1965}. The RIME computed by the GPU kernels is
unit tested against CPU numpy \citep{numpy} code, accelerated
with the numexpr library \citep{numexpr}.

\section{Results}

\noindent
We tested our code on a high-performance computer configured with a dual
Intel Xeon(R) hexacore E5--2620v2 3.00GHz CPU, 128GB of memory
and a NVIDIA K40 Tesla card.

\begin{table}
\begin{center}
\begin{tabular}{|l|l|r|r|r|r|r|}
\hline
Antennas & Baselines & Montblanc (s) & \textsc{OSKAR} (s) & Ratio & \textsc{MeqTrees} (s) & Ratio \\
\hline
\multicolumn{7}{|c|}{float} \\
\hline
7 & 21 & 0.0042 & 0.3980 & 93.51 & -- & -- \\
14 & 91 & 0.0136 & 0.4314 & 31.67 & -- & -- \\
27 & 351 & 0.0435 & 0.6246 &  14.33 & -- & -- \\
64 & 2016 & 0.2320 & 1.7840 & 7.69 & -- & -- \\
128 & 8128 & 0.7007 &  5.8305 & 8.32 & -- & -- \\
192 & 18336 & 1.5730 & 12.4372  & 7.90 & -- & -- \\
\hline
\multicolumn{7}{|c|}{double} \\
\hline
7 & 21 & 0.0073 & 0.4504 & 61.28 &  1.8183 & 247.39 \\
14 & 91 & 0.0215 & 0.6164 & 28.62 &  5.5171 & 256.13 \\
27 & 351 & 0.0703 & 1.1169 & 15.87 & 17.3753  & 246.94 \\
64 & 2016 & 0.3668 & 4.5315 & 12.35 &  96.0943 &  261.98 \\
128 & 8128 &  0.9113 & 16.8675 & 18.51 & --  & -- \\
192 & 18336 & 2.3540 & 37.4096 & 15.89  & -- & -- \\
\hline
\end{tabular}
\end{center}
\caption{Timings for different problem sizes. This table
shows the time taken to compute the RIME for varying
numbers of antennas and baselines, as well as the speedup
that Montblanc achieves vs \textsc{OSKAR} and \textsc{MeqTrees}.
64  channels, 100 timesteps, 50 point and 50 Gaussian sources
were used  in all cases.}
\label{tbl:timings}
\end{table}

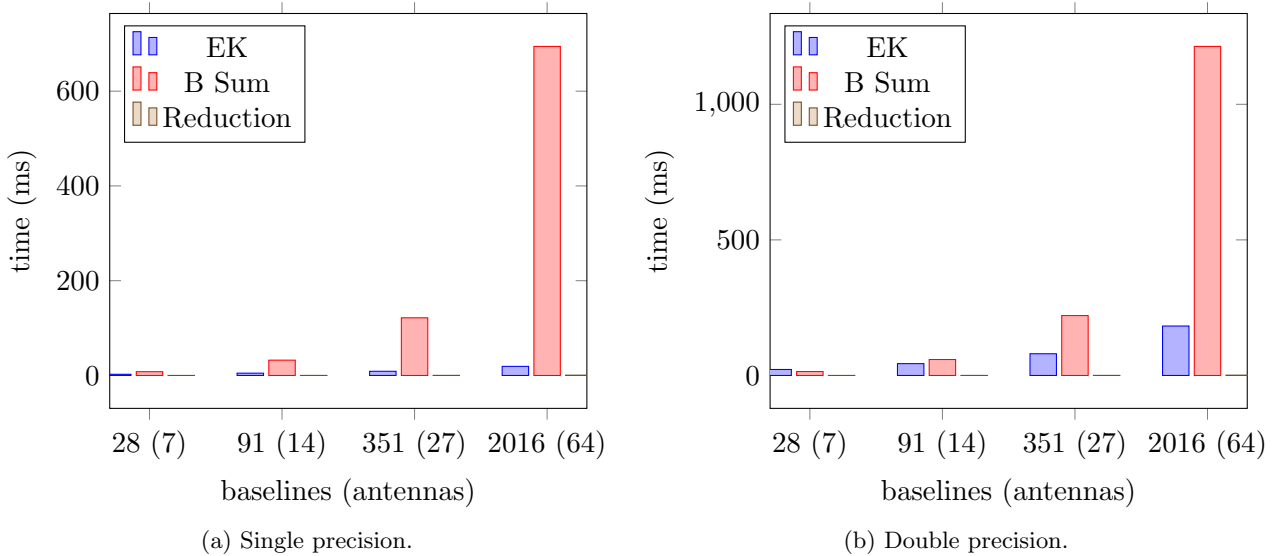
\begin{figure}
\begin{center}
\subfloat[Single precision.] {
\begin{tikzpicture}
\begin{axis}[
    %ymode=log,
    width=0.48\textwidth,
    xtick={0,1,2,3},
    xticklabels={28 (7),91 (14) ,351 (27),2016 (64)},
    xlabel={baselines (antennas)},
    ylabel=time (ms),
    ybar,
    legend entries={EK, B Sum, Reduction},
    legend pos=north west]
% 90 180 349 833
\addplot coordinates {(0,2.5275520) (1,4.8537181)  (2,8.8371643) (3,19.2224833)};           % EK
\addplot coordinates {(0,7.9459840) (1,32.3258043)  (2,121.7097637) (3,694.4441970)};       % B Sum
\addplot coordinates {(0,0.0183261) (1,0.0463901) (2,0.150641) (3,0.5636233)};                    % Reduction
\end{axis}
\end{tikzpicture}
\label{fig:kernel_times_float}
}
\subfloat[Double precision.] {
\begin{tikzpicture}
\begin{axis}[
    %ymode=log,
    width=0.48\textwidth,
    xtick={0,1,2,3},
    xticklabels={28 (7),91 (14) ,351 (27),2016 (64)},
    xlabel={baselines (antennas)},
    ylabel=time (ms),
    ybar,
    legend entries={EK, B Sum, Reduction},
    legend pos=north west]
]
\addplot coordinates {(0,22.1006523) (1,43.6285986)  (2,80.1101257) (3,182.3552981)};
\addplot coordinates {(0,14.4905846) (1,58.8435488)  (2,220.8617127) (3,1213.5170000)};
\addplot coordinates {(0,0.0248984) (1,0.0676971) (2,0.1859716) (3,1.0514116)};
\end{axis}
\end{tikzpicture}
\label{fig:kernel_times_double}
}
\end{center}
\caption{Single and double-precision kernel running times for different baseline
sizes. The B Sum kernel dominates computational cost as the number of antennas, and
by implication, baselines increase.}
\label{fig:kernel_times}
\end{figure}

\subsection{Kernel running times and computational complexity}
\label{sec:kernel_times_comp_complexity}

\noindent
The running times for our kernels are shown in
Figure~\ref{fig:kernel_times} for different baselines.
With a computational complexity of
$O(\textrm{log}_2 (\mathtt{ntime \times nbl \times nchan}))$, the
reduction kernel contributes minimally to computational cost.
By contrast, the B Sum kernel consumes most of the cost once
the number of baselines starts to scale. This is because the computational
complexity of the different kernels differ: $O(\mathtt{ntime \times na
  \times nsrc \times nchan})$ and $O(\mathtt{ntime \times nbl \times
  nsrc \times nchan})$ for the EK and B Sum kernels respectively. Due
to the quadratic relation between the number of baselines and the
number of antennas, the B Sum kernel will, in general, consume
quadratically more time. Therefore, Montblanc takes on the
computational complexity of the B Sum kernel, i.e.
$O(\mathtt{ntime \times  nbl \times nsrc \times nchan})$.

These figures support our decision to separate
expensive, per-antenna computation into the EK kernel while
assigning the multiplication and summation steps to the B Sum kernel:
increasing the expense of the dominant complexity factor is a poor design choice.
For example, the EK kernel is slower for the 7-antenna double-precision case,
but the B Sum kernel rapidly outstrips it on the 64-antenna case.

Folding EK and B Sum kernels into a monolothic kernel operating over the
{\tt  ntime}, {\tt na} and {\tt nchan} dimensions is an interesting
proposition: the EK kernel's compute capabilities could be used to hide
the latency of the B sum kernel even further by
 computation of expensive per-antenna values.
These values could be output to DRAM,
and then multiplied and summed based on a
triangular number pattern.
However, this is not
currently possible due to CUDA's lack of
block-level synchronisation.

Additionally, combining both the calculation, multiplication and sum
of per-antenna terms into a monolithic kernel
increases register usage and decreases occupancy.
The double-precision B sum kernel already uses 63 registers
and has a 50\% occupancy. Further reductions in occupancy
would reduce the solution's parallelism.

\subsection{Comparison with \textsc{OSKAR}}

\begin{figure}
\begin{center}
\subfloat[\textsc{OSKAR}.] {
\begin{tikzpicture}
\begin{axis}[
    width=0.48\textwidth,
    xtick={2,3,4,5},
    xticklabels={27,64,128,192},
    xlabel=antenna,
    ylabel=Speedup,
    legend entries={float, double},
    ybar]
% OSKAR float
\addplot coordinates {
%    (0,93.51)
%    (1,31.67)
    (2,14.33)
    (3,7.69)
    (4,8.32)
    (5,7.90)
};
% OSKAR double
\addplot coordinates {
%    (0,61.28)
%    (1,28.62)
    (2,15.87)
    (3,12.35)
    (4,18.51)
    (5,15.89)
};
\end{axis}
\end{tikzpicture}
\label{fig:mb_oskar_ratios}
}
\subfloat[\textsc{MeqTrees}.] {
\begin{tikzpicture}
\begin{axis}[
    width=0.48\textwidth,
    xtick={0,1,2,3},
    xticklabels={7,14,27,64},
    xlabel=antenna,
    ylabel=Speedup,
    ymin=0, ymax=300,
    ybar]
\addplot coordinates {
    (0,247) % 0.007350s vs 1.818300s
    (1,256) % 0.021540s vs 5.517110s
    (2,247) % 0.070362s vs 17.375343s
    (3,262) % 0.366803s vs 96.094316s
};
\end{axis}
\end{tikzpicture}
\label{fig:cpu_gpu_ratios}
}
\caption{The ratio, or speedup, of Montblanc's GPU
  implementation on a NVIDIA K40 Kepler versus \textsc{MeqTrees}
  and \textsc{OSKAR's} RIME implementation for varying antenna numbers.
  \textsc{MeqTrees} executed on a dual hexacore E5--2620v2 system, while
  \textsc{OSKAR} executed on the K40. 64  channels, 100 timesteps,
  50 point and 50 Gaussian sources were used  in all cases.}
\end{center}
\end{figure}
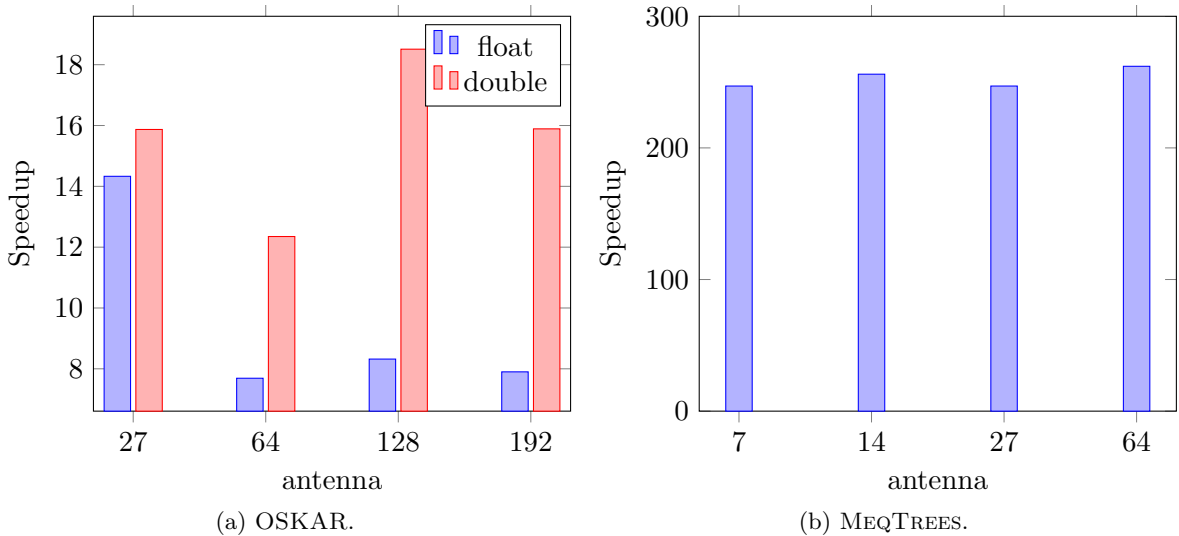

We performed a performance comparison between
Montblanc and OSKAR's interferometer simulator.
OSKAR supports parts of the RIME on the CPU and the
GPU. In the OSKAR output below, it can be seen that
horizon clipping and the Jones E term, both implemented
on the CPU, together consume almost 80\%
of simulation time.

\begin{Verbatim}[samepage=true,commandchars=\\\{\},codes={\catcode`$=3\catcode`_=8}]
=|== Simulation completed in 16.345 sec.
 |
 |  2.6% Chunk copy & initialise.
 | 69.1% Horizon clip.
 |  6.3% Jones R.
 | 10.3% Jones E.
 |  0.4% Jones K.
 |  1.0% Jones join.
 |  5.5% Jones correlate.
 |
 | + Writing Measurement Set: 'VLA27.MS'
 |
=|== Run complete.
\end{Verbatim}

\begin{figure}
\begin{center}
\begin{tikzpicture}
\begin{axis}[
    width=0.75\textwidth,
    xlabel={\tt ntime x nbl x nsrc x nchan},
    ylabel={time (s)},
    legend entries={baseline,channel,time,point source,gaussian source},
    legend pos=south east]
% na
\addplot coordinates {
    (672000000,0.008685)
    (2912000000,0.025978)
    (11232000000,0.100713)
%    (64512000000,0.545635)
};
% nchan
\addplot coordinates {
    (672000000,0.007796)
    (1344000000,0.016824)
    (2688000000,0.031869)
    (5376000000,0.062046)
};
% ntime
\addplot coordinates {
    (672000000,0.008644)
    (2016000000,0.024573)
    (4032000000,0.048224)
    (6720000000,0.070540)
};
% npsrc
\addplot coordinates {
    (672000000,0.008850)
    (4032000000,0.020963)
    (8064000000,0.037433)
    (13440000000,0.057506)
};
% ngsrc
\addplot coordinates {
    (672000000,0.007973)
    (4032000000,0.025601)
    (8064000000,0.045634)
    (13440000000,0.071312)
};
\end{axis}
\end{tikzpicture}
\caption{Rime Dimension Variation.
This graph shows the effect of varying
each RIME dimension on the execution
time. Adding timesteps and channels is twice
as expensive as adding sources.}
\label{fig:dim_variation}
\end{center}
\end{figure}
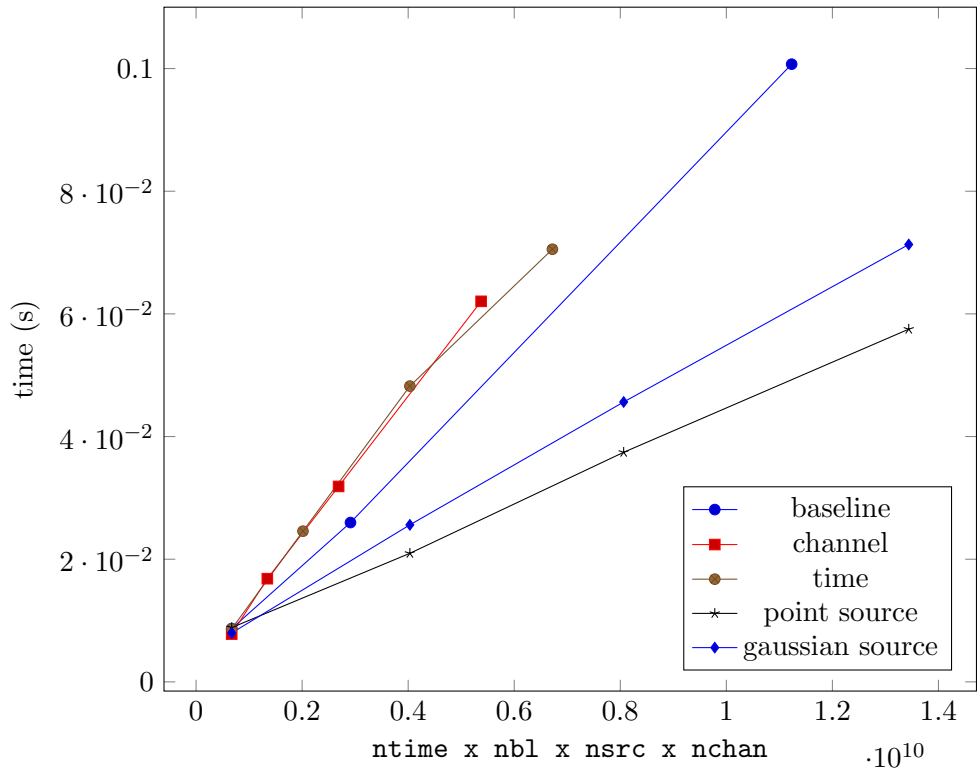

Therefore in our comparison, we have chosen
to compare only those parts of the RIME that are present
on the GPU, namely the Jones K kernel, Jones join kernel
and the Jones correlate kernel. Thus we estimated OSKAR's
performance for the above case as follows:
$16.345s \times (0.4\% + 1.0\% + 5.5\%) = 1.127805s$.

Note that Montblanc's GPU timing includes both a
E term, computation of the $\chi^2$ terms and reduction
to a single $\chi^2$ value. Montblanc's E term is however,
simpler than OSKAR's. As such, it is difficult to perform a
direct comparison, but the supplied figures provide
an indication of relative performance.

Table \ref{tbl:timings} and Figure \ref{fig:mb_oskar_ratios}
show that Montblanc is much faster than OSKAR for
smaller antenna numbers, but this advantage drops
when the number of antennas (and baselines) is large.
In these cases Montblanc is at least 7.7 and
12 times faster than OSKAR for single and double-precision
floating point respectively.

\subsection{Comparison with \textsc{MeqTrees}}

\noindent
We performed a comparison of Montblanc's GPU implementation with
\textsc{MeqTrees}. However, the maximum theoretical performance
of the B Sum kernel on the dual Xeon system is lower than on the K40.
If the B Sum kernel, which dominates the computation for large problem sizes,
were to be implemented on the CPU, it would, at most be able to
achieve $1.75 \times 51.2 = 89.6$ GFLOPS/s. This is 44\% of the Xeon's
theoretical peak performance, and $5.625$
times less than the kernel's theoretical peak of $504$ GFLOPS/s on the K40
(Figure \ref{fig:roofline}.

The speed-ups achieved are shown in Figure \ref{fig:cpu_gpu_ratios}. Channels,
timesteps, point and Gaussian sources were held constant at dimensions
of 64, 100, 50 and 50 respectively. The number of antennas was
varied between 7, 14, 27 and 64, corresponding to the KAT7, Westerbork,
VLA and MeerKAT telescopes. \textsc{MeqTrees} was configured to use all
12 cores available to the system. It can be seen that Montblanc is around
250X faster than \textsc{MeqTrees}, due to the disparity between
CPU and GPU architectures.

\subsection{RIME Dimension Variation}

\noindent
We performed analysis of our RIME solution to investigate the expense
of increasing each dimension. This involved varying each dimension
from a base configuration of 7 baselines, 64 channels, 100 timesteps,
50 point sources and 50 Gaussian sources. Figure
\ref{fig:dim_variation} shows the total number of elements,
$\mathtt{ntime \times nbl \times nsrc \times nchan}$ versus the time
taken to solve the RIME.

It can be seen that increases in each dimension result in linear
performance scaling. The numbers of timesteps and channels contribute
most to the expense of the solution, followed by the baseline
dimension. Gaussian and point sources are the least expensive
dimensions to increase -- roughly half as expensive as
timesteps and channels. Point sources are marginally less
expensive than Gaussians.

We expect that the {\tt ntime} and {\tt nsrc} dimensions will vary
most for different problems, since channels and baselines will likely
remain static for individual interferometers.

\section{Conclusion}

\noindent
In this paper we presented Montblanc, a GPU-accelerated framework
for solving the RIME in order to accelerate the BIRO technique.
Using NVIDIA's CUDA architecture,
it computes the RIME  and differences the resulting model visibilities
with those observed by an interferometer to produce multiple $\chi^2$
likelihood values. These values are used to drive the Bayesian inference process.

At present, it supports point and Gaussian sources, time-varying source
brightness and both single and double-precision solutions. It is
architected to allow addition of other source morphologies, such as $\beta$
profiles and the computation can be subdivided in order to fit large
problem sizes within a single GPUs memory. While only single GPUs are
currently supported, this subdivision makes multi-GPU and clustered
solutions to the RIME viable.

Expensive, per-antenna terms are computed in an EK kernel, and
output to the GPU's main memory. These per-antenna terms are read
in by a B Sum kernel, and combined to form per-baseline terms. Model
visibilities are computed from these terms and a $\chi^2$ value that
is used by BIRO in the Bayesian inference process.

We have analysed our CUDA kernels, showing that
one kernel dominates the run-time. Theoretical analysis suggests that
the arithmetic intensity of this kernel is 1.75 FLOPS/byte, indicating
a {\itshape memory} bound algorithm. By contrast, profiling indicates
a balanced kernel that is neither {\itshape compute} nor {\itshape memory}
bound. This is due to the fact that much of the problem is retained
in Kepler's L1 and L2 cache, avoiding expensive trips to the GPU DRAM.
The computational complexity of Montblanc is
$O(\mathtt{ntime \times nbl \times nsrc \times nchan})$.

We compared Montblanc's performance on an NVIDIA K40 Kepler GPU
with MeqTrees execution on a dual hexacore E5--2620v2 Xeon system.
For a problem size of 64 antennas, 100 timesteps, 64 channels, 50
point and 50 Gaussian sources, Montblanc is around 250 times faster.
We also compared it against parts of the OSKAR simulator's CUDA RIME
pipeline, and found that it was at least 7.7 and 12 times faster for
single and double-precision floating point cases, respectively.

Montblanc is implemented as a Python package, and is designed for
the addition of new source profiles. It is not limited to use for BIRO;
it computes visibilities and can be used as a fast simulator.
Users may implement their own
RIME implementation with a custom pipeline, if required.
With the future addition of more general DIE and DDE matrices,
it could also be used as a calibrator.

\subsection{Future Work}

\noindent
As BIRO currently operates on fully calibrated data, it was not
immediately necessary to include evaluation of the
$G_{ps}$ (DIE) matrices. Additionally, analytic expressions were
substituted for the $E_{ps}$ (DDE) terms. In future work, we intend
to provide full support for these terms as matrices. In the case of
the DIE matrices, this will allow Montblanc to perform sky modeling
and calibration simultaneously.

The high parallelism inherent in the calculation of the BIRO RIME
and $\chi^2$ calculation also make the problem amenable to
subdivision. We aim to provide support for multiple GPUs and
distribution of the computation amongst multiple High Performance
Computing nodes. This will necessarily involve implementing an out-of-core
solution for BIRO, and providing solutions for the
data-transfer bottlenecks inherent to this solution.

\section{Acknowledgments}

\noindent The authors wish to acknowledge the following for their support and advice: \\
\, \textbullet \, Marzia Rivi for corrections to the manuscript. \\
\, \textbullet \, Stefan van der Walt for numpy and numexpr advice. \\
\, \textbullet \, Bruce Merry and Ben Hugo for CUDA/PyCUDA conversations. \\
\, \textbullet \, Michael Clark from NVIDIA. \\
\, \textbullet \, Andrew Lewis and Tim Carr for help compiling on hex. \\
\, \textbullet \, Computations were performed using facilities provided by the University of Cape Town's ICTS High Performance Computing team: \url{http://hpc.uct.ac.za} \\
\, \textbullet \, Simon Perkins acknowledges SKA South Africa for a postdoctoral research fellowship. \\
\, \textbullet \, Jonathan Zwart acknowledges SKA South Africa for a postdoctoral research fellowship. \\
\, \textbullet \, Iniyan Natarajan acknowledges the HPC for Radio Astronomy Programme. \\
\, \textbullet \, Oleg Smirnov's research is supported by the South African Research Chairs
Initiative of the Department of Science and Technology and National Research Foundation. \\

\bibliography{montblanc}
\bibliographystyle{elsarticle-harv}

\end{document}